\documentclass[fleqn,usenatbib]{mnras}

\usepackage{amsmath}
\IfFileExists{newtxtext.sty}{\usepackage{newtxtext,newtxmath}}{\usepackage{txfonts}}
\usepackage{booktabs}
\usepackage{makecell}
\usepackage{subfig}

\usepackage{xcolor}
\definecolor{revisionblue}{RGB}{35,95,200}
\newif\ifshowrevisions
\showrevisionstrue

\usepackage{placeins}
\usepackage{lastpage}
\usepackage[T1]{fontenc}

\DeclareRobustCommand{\VAN}[3]{#2}
\let\VANthebibliography\thebibliography
\def\thebibliography{\DeclareRobustCommand{\VAN}[3]{##3}\VANthebibliography}

\usepackage{graphicx}	
\usepackage{amsmath}	

\title[DM mapping in Solar Neighbourhood]{Towards three-dimensional dark matter mapping in the extended solar neighbourhood with Gaia Data Release 3 and deep learning}

\author[Y. Kim et al.]{
Yigon Kim,$^{1}$\thanks{E-mail: yigon28@snu.ac.kr}
Ho Seong Hwang,$^{1,2,3}$\thanks{E-mail: hhwang@astro.snu.ac.kr}
Sungwook E. Hong,$^{4,5}$
Donghui Jeong,$^{6,7}$
Mingyeong Yang$^{4,5}$
\newauthor
and Ji-hoon Kim$^{2,3,8}$
\\
$^{1}$Astronomy Program, Department of Physics and Astronomy, Seoul National University, 1 Gwanak-ro, Gwanak-gu, Seoul 08826, Republic of Korea\\
$^{2}$SNU Astronomy Research Center, Seoul National University, 1 Gwanak-ro, Gwanak-gu, Seoul 08826, Republic of Korea\\
$^{3}$Institute for Data Innovation in Science, Seoul National University, Seoul 08826, Republic of Korea\\
$^{4}$Korea Astronomy and Space Science Institute (KASI), Daejeon 34055, Republic of Korea\\
$^{5}$Astronomy campus, University of Science and Technology, Daejeon 34055, Republic of Korea\\
$^{6}$Department of Astronomy and Astrophysics and Institute for Gravitation and the Cosmos, The Pennsylvania State University, University Park, PA 16802, USA\\
$^{7}$School of Physics, Korea Institute for Advanced Study, 85 Hoegiro, Dongdaemun-gu, Seoul 02455, Republic of Korea\\
$^{8}$Center for Theoretical Physics, Department of Physics and Astronomy, Seoul National University, Seoul 08826, Republic of Korea
}

\date{Accepted XXX. Received YYY; in original form ZZZ}

\pubyear{\the\year{}}

\begin{document}
\label{firstpage}
\pagerange{\pageref{firstpage}--\pageref{LastPage}}
\maketitle

\begin{abstract}
We investigate a method for inferring the local dark matter (DM) density in the solar neighbourhood using a deep learning framework that learns the mapping between stellar kinematics and the underlying DM density field from cosmological simulations. To this end, we construct a training set from $7$ Milky Way–like galaxies in the FIRE-2 simulation suite. We use three-dimensional stellar kinematic components, consisting of the tangential velocities along the right ascension and declination directions ($v_{\alpha*}$ and $v_{\delta}$) derived from parallaxes and proper motions, together with radial velocity, as input features to predict the three-dimensional DM density field. Two distinct architectures, VNet \citep{Milletari2016} and ViT-VNet \citep{Chen2021}, are trained on identical datasets. Both models reproduce the large-scale DM density gradient, whereas small-scale fluctuations are systematically smoothed. Applying the trained models to Gaia DR3 data yields our best estimates of the local DM density: $0.80^{+0.10}_{-0.10}$ and $1.12^{+0.15}_{-0.13}\,\mathrm{GeV}\,\mathrm{cm}^{-3}$ from VNet and ViT-VNet, respectively. The quoted uncertainties reflect the inference scatter across the training set. We further examine systematics from model dependence and the simulation--Gaia domain mismatch, providing an initial step towards a framework for reconstructing local DM maps from future multi-epoch astrometric measurements, including those expected from Gaia DR4.
\end{abstract}




\section{Introduction} \label{sec:intro}

Dark matter (DM) comprises roughly five times more energy density than baryonic matter in the Universe \citep{WMAP2009, Planck2018}, resulting in a dominant contribution to the gravitational potential. Such influence is present across the whole Universe and underlies the great success of modern cosmology \citep{Zeldovich1970, Press1974, White1978, Bardeen1983, White1991, Sheth1999, Sheth2001}. Consequently, studying the spatial distribution of DM is necessary for a complete understanding of the status and history of the luminous components of the Universe.

Galaxies are no exception; the mass, kinematics, and evolutionary history of galaxies are deeply related to the DM distribution, forming a DM halo that encloses the whole galaxy \citep{Navarro1996, Navarro1997}. However, there are tensions between the traditionally favoured cold dark matter (CDM) model and observations \citep{Weinberg2015, Bullock2017}, e.g., the cored density profile at the centre of DM haloes (known as the core–cusp problem; \citealt{deBlok2010, Genina2018, DelPopolo2021}) and the lower-than-expected halo masses of dwarf galaxies (the too-big-to-fail problem; \citealt{BoylanKolchin2011, BoylanKolchin2012, Papastergis2015}), motivating alternative DM models \citep{Carlson1992, Dodelson1994, Shi1999, Spergel2000, Maccio2012, Rocha2013, TulinYu2018}.

Consequently, various DM models can be tested on small scales, such as the central region of DM haloes, subhaloes, and the local phase-space distribution. This compels us to finely resolve the spatial distribution of DM haloes, beyond merely measuring the density profile and total mass of the halo itself. Integral field spectroscopic observations provide spatially resolved kinematics of galaxies for this requirement \citep{MUSE, CALIFA, MANGA, SAMI}. But, achieving such a high-resolution map of DM remains challenging due to line-of-sight projection effects, distance uncertainties, and the limited spatial resolution of the luminous components of galaxies.

The only galaxy for which we can measure full six-dimensional kinematics of individual stars is our Milky Way (MW). The Hipparcos satellite first provided precise all-sky astrometric data for MW stars \citep{Hipparscos, Tycho2}, and Gaia has succeeded it by delivering even richer astrometric measurements \citep{Gaia}. The third data release (DR3; \citealt{GaiaDR3_summary}) contains measurements for $\sim1.8$ billion detected sources, among which more than $34$ million have radial velocity (RV) measurements and therefore provide full six-dimensional phase-space information. As stars orbit within the MW under its gravitational potential, they become excellent tracers for constructing the density map. The MW is therefore a natural laboratory for mapping the high-resolution three-dimensional DM structure.

Such ideas, combined with precise astrometry, have produced valuable measurements of the local DM density through analyses of the rotation velocities \citep{Xue2008, Boby2012, McMillan2017, deSalas2019, Cantun2020, Ou2024}, distribution functions \citep{Kuijken1989, Kuijken1991, Binney2015, Cole2018}, and the motions of halo objects \citep{Watkins2010, Posti2019} and others \citep{Garbari2012, Lim2025}. See \citet{Read2014, deSalasWidmark2021} for a review. While these efforts necessitate testing the consistency between DM halo models and galactic dynamics, they rely on certain assumptions, such as axisymmetry and virial equilibrium of the halo. Since the gravitational potential is directly related to the matter density through the Poisson equation, the stellar acceleration actually is the direct probe of the underlying mass distribution. However, resolved measurements of stellar acceleration for individual stars in time series over large volumes are currently not available. Thus, local density measurements without retrieving an actual map of the DM distribution inherently encapsulate uncertainties.

Therefore, achieving a more accurate characterisation of DM and a deeper understanding of the dynamics of the luminous components of the MW necessitates the construction of a three-dimensional map of the DM distribution. In addressing similar challenges on cosmological scales, recent studies have attempted to infer the DM distribution from the luminous component through dynamical and statistical approaches \citep{Yoo2025, SKim2026}. Among these, several works have introduced deep learning (DL) techniques trained to reconstruct the large-scale DM structure from the phase-space information of galaxies \citep{Hong2021, Ganeshaiah2023, deAndres2024, Dupuy2026}, thereby inferring the relation between the kinematics of the luminous component and the underlying DM density.

In this study, we extend these efforts to the internal region of the MW by employing a DL-based approach to reconstruct the three-dimensional distribution of DM from the six-dimensional phase-space information of nearby stars provided by Gaia DR3. Our ultimate goal is to develop and evaluate a new method for mapping the DM density in the solar neighbourhood and by doing so, infer the local DM density of the MW.

This paper is organised as follows. The overall methodology is described in Sec.~\ref{sec:method}. In Sec.~\ref{sec:method_DL}, we introduce the deep learning architectures employed in this study, and the training strategy is presented in Sec.~\ref{sec:DL_train}. The datasets are described in Sec.~\ref{sec:data}, including the simulation data for training (Sec.~\ref{sec:data_sim}) and the Gaia data used to infer the DM density field in the solar neighbourhood (Sec.~\ref{sec:data_gaia}). Additional details regarding data subsampling are provided in Sec.~\ref{sec:method_sel}. The results are presented in Sec.~\ref{sec:result}, followed by a discussion in Sec.~\ref{sec:discuss}. Finally, we summarise this study in Sec.~\ref{sec:concl}.

Throughout this paper, we adopt the following MW parameters from \citet{BlandHawthorn2016}: $R_{200\mathrm{m}}=320\,\mathrm{kpc}$, defined as the radius enclosing a mean density $200$ times the cosmological mean matter density, and $M_{200\mathrm{m}}=1.1\times10^{12}\,\mathrm{M}_{\odot}$, defined as the mass within this radius; and the scale radius of the Navarro–Frenk–White (NFW) density profile \citep{Navarro1997}, $R_s = 25\,\mathrm{kpc}$.

\section{Method} \label{sec:method}

\subsection{Deep learning architecture} \label{sec:method_DL}

Selecting an appropriate architecture is a prerequisite for training a DL model tailored to a specific scientific objective. In our case, the task can be formulated as a volumetric image generation problem. Within this framework, we employ two architectures: a VNet and a hybrid Vision Transformer–VNet (ViT-VNet) after considering computational cost under our resources. All models were trained using two NVIDIA L40S GPUs, each equipped with 48GB of Video RAM.

\begin{figure}
    \centering
    \includegraphics[width=\columnwidth]{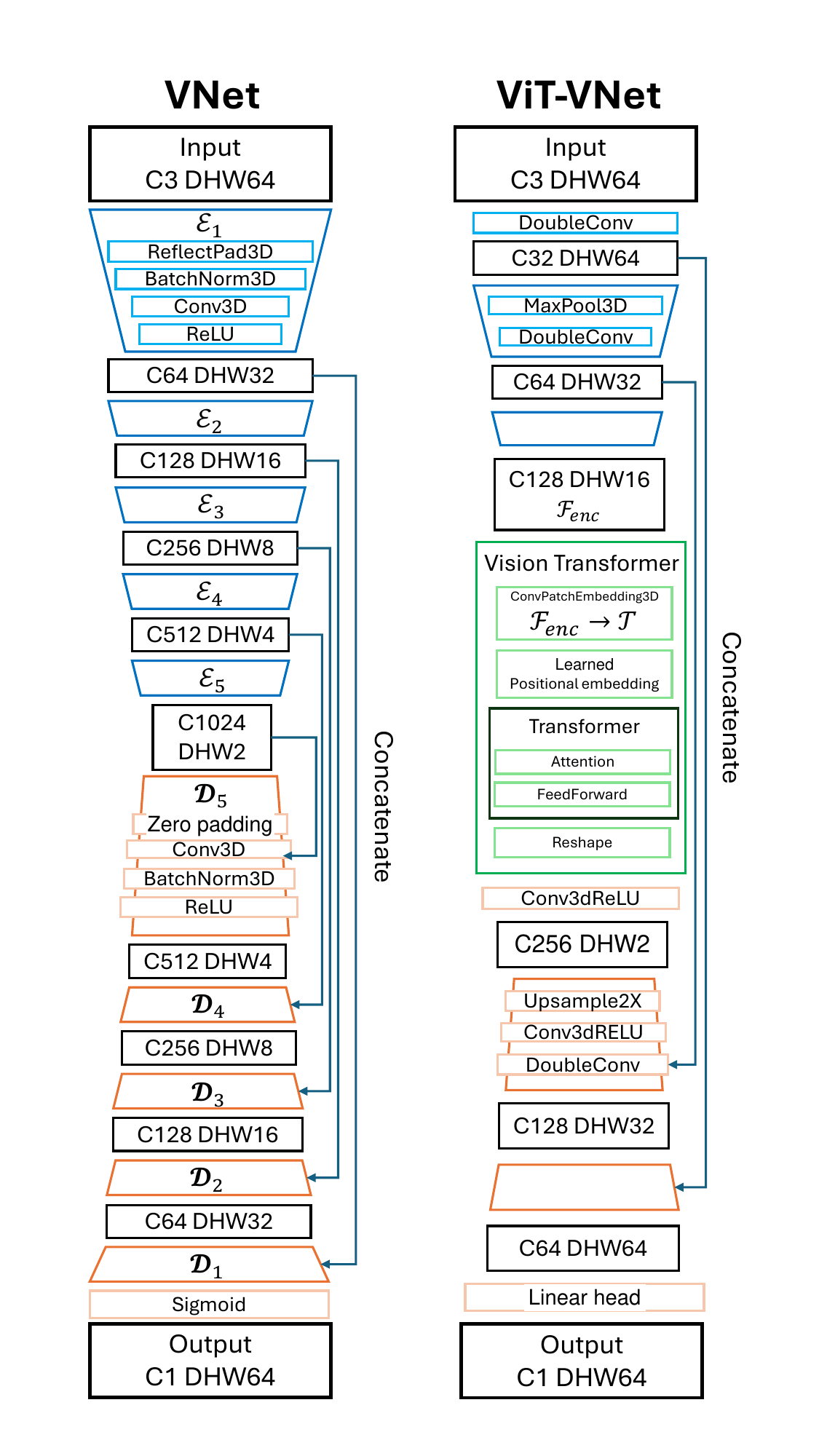}
    \caption{Schematic diagram of the VNet and ViT-VNet models. The notation $\mathrm{C}x$ and $\mathrm{DHW}y$ denote the tensor shape, where $x$ indicates the number of channels and $y$ represents the cubic spatial dimension along each axis.}
    \label{fig:DL_architect}
\end{figure}

\subsubsection{VNet} \label{sec:DL_vnet}

VNet is a three-dimensional extension of convolutional encoder–decoder architectures originally developed for volumetric segmentation tasks \citep{Milletari2016}. It is one of a variety of convolutional neural networks (CNNs) designed to process structured grid data through hierarchical feature extraction. The architecture consists of an encoding path and a decoding path, connected through skip connections.

Each encoding stage progressively reduces the spatial resolution while increasing the number of feature channels, enabling the network to capture increasingly abstract representations of the input. The decoding path subsequently restores spatial resolution by combining high-level semantic features with fine-grained information from earlier layers via skip connections. This design makes VNet particularly suitable for dense prediction tasks in three-dimensional voxelised domains.

We tailored this architecture to reconstruct the voxelised DM density field from preprocessed stellar phase-space information, following the general framework introduced by \citet{Hong2021} for large-scale DM structure reconstruction. The input data consist of 
$C$ volumetric channels with spatial resolution $D \times H \times W$, each representing a distinct phase-space feature. The network output is a single-channel volumetric map of identical spatial dimensions, corresponding to the reconstructed DM density field. Accordingly, we denote the input and output tensors as
\begin{equation}
    \mathbf{X} \in \mathbb{R}^{C \times D \times H \times W}\,\, , \,\,\mathbf{Y} \in \mathbb{R}^{D \times H \times W}.
\end{equation}

The architecture first encodes the input features using three-dimensional convolutional filters of size $5\times5\times5$. The convolution is applied with a stride of 2, thereby reducing the spatial resolution of the volume by a factor of two while increasing the number of feature channels. This design enables hierarchical feature extraction while preserving essential structural information from the original input.

Prior to the convolution, reflection padding of two voxels is applied along each spatial boundary to ensure consistent treatment of edge regions and to avoid artificial boundary truncation during convolution. Batch normalisation is then applied, followed by the convolution operation. The resulting feature maps are subsequently passed through an activation function; in this study, we adopt the Rectified Linear Unit (ReLU). Thus, a single encoding layer can be expressed as an operator $\mathcal{E}$.
\begin{equation}
    \mathcal{E} \left( \mathbf{X} \right) = \mathrm{ReLU} \left( \mathrm{Conv3D_{k=5,s=2}} \left( BN \left( \mathcal{P}_{\mathrm{reflect}} \left( \mathbf{X} \right) \right) \right) \right)
\end{equation}

The decoding path progressively restores the spatial resolution of the feature maps while integrating multiscale information from the encoding path. At each decoding stage, the upsampled feature map is concatenated with the corresponding feature map from the encoder via a skip connection. This operation preserves fine-grained spatial information that would otherwise be lost during downsampling. 

The concatenated feature maps are then processed by a three-dimensional convolution with kernel size $3\times3\times3$, followed by batch normalisation and a nonlinear activation. For intermediate layers, we adopt the ReLU, while a sigmoid activation is applied at the final layer to produce a normalised density prediction while bounding its range for stable training and physical interpretability. We express the decoding layer as $\mathcal{D}$:
\begin{equation}
    \begin{aligned}
        \mathcal{D} \left( \mathbf{X}', \mathbf{F}_{\mathrm{skip}} \right)
        &= \phi \Bigl( \mathrm{BN} \Bigl( \mathrm{Conv3D}_{k=3,s=1} \Bigl( \\
        &\quad \mathcal{P}_{\mathrm{zero}} \Bigl(
            \mathrm{Concat} \bigl[
                \mathcal{U}_{2} \left( \mathbf{X}' \right),
                \mathbf{F}_{\mathrm{skip}}
            \bigr]
        \Bigr) \Bigr) \Bigr) \Bigr)
    \end{aligned}
\end{equation}

\noindent where $\mathcal{U}_{2}$ denotes spatial upsampling by a factor of two along each axis, and $\mathbf{F}_{\mathrm{skip}}$ represents the corresponding feature map from the encoder. The upsampled feature map is concatenated with the encoder feature map along the channel dimension before convolution. The decoder convolution uses zero padding of one voxel. The activation function $\phi$ is Sigmoid for the final layer and ReLU for the intermediate layers. As a result, the whole architecture with $L$ layers can be expressed as Eq.~4. We set $L=5$ throughout training for this study, as schematised in the left panel of Fig.~\ref{fig:DL_architect}.

\begin{equation}
    \mathbf{Y} = \mathcal{D}_1 \circ \cdots \ \circ \mathcal{D}_{L} \circ \mathcal{E}_{L} \circ \cdots \ \circ \mathcal{E}_1 \left( \mathbf{X} \right)
\end{equation}

This VNet implementation contains $115\,273\,556$ trainable parameters.

\subsubsection{Hybrid Vision Transformer-VNet} \label{sec:DL_vit}

Transformers were originally developed to model long-range dependencies in natural language processing. Variants of this architecture have since been adapted for image analysis by dividing an image into patches, treating them as tokens, and learning global contextual relationships through self-attention mechanisms. This adaptation is commonly referred to as the Vision Transformer (ViT; \citealt{Dosovitskiy2021}).

However, because ViT relies primarily on attention mechanisms without explicit convolutional inductive biases, its ability to capture fine-grained local details may be limited. In addition, the computational and memory costs increase rapidly as the spatial resolution of the input becomes finer. To address these limitations, hybrid architectures have been proposed that combine transformer blocks with convolutional networks. In particular, ViT-VNet integrates a ViT module with convolution-based encoder–decoder structures for volumetric medical image registration \citep{Chen2021}, effectively incorporating both long-range contextual information and local structural details. 

Motivated by this design, we adopt a similar hybrid structure in which a ViT module is embedded within a volumetric convolutional framework for three-dimensional DM density reconstruction. The network consists of three main components: a CNN encoder, a Transformer bottleneck operating on volumetric patches, and a convolutional decoder with skip connections.

Given an input volume $\mathbf{X} \in \mathbb{R}^{C \times D \times H \times W}$, a lightweight 3D convolutional encoder first extracts multi-scale feature maps. The deepest feature representation, denoted as $\mathbf{F}_{\mathrm{enc}}$, is used for patch embedding, while intermediate feature maps are preserved for skip connections. Patch embedding is implemented via a 3D convolutional projection, 
\begin{equation}
    \mathbf{T} = \mathrm{Flatten}\left( \mathrm{Conv3D}_{\mathrm{patch}} \left( \mathbf{F}_{\mathrm{enc}} \right) \right)
\end{equation}

\noindent which converts the volumetric feature map into a sequence of $N$ tokens, $\mathbf{T} \in \mathbb{R}^{N \times d}$ where $d$ is the embedding dimension. To preserve spatial information, a learned positional embedding is added. The token sequence is then processed by stacked Transformer encoder layers, enabling global interactions across the entire volume through self-attention mechanisms \citep{Vaswani2017}. After transformation, the tokens are reshaped back into a volumetric feature map. For a $64^3$ input volume, we apply $8\times8\times8$ patches with the same stride to the deepest encoder feature map. The Transformer uses an embedding dimension of $256$ and comprises three layers with eight attention heads each.

In the last stage, the transformed feature map is progressively upsampled through a convolutional decoder. At each resolution level, feature maps from the encoder are concatenated via skip connections. The final output layer with a linear activation function produces a voxelwise scalar field, $\mathbf{Y} \in \mathbb{R}^{D \times H \times W}$, representing the DM density field. This configuration contains $23\,214\,172$ trainable parameters, including the positional embedding. We visualised the whole architecture in the right side of Fig.~\ref{fig:DL_architect}.

\subsection{Training} \label{sec:DL_train}

Both architectures are trained in a supervised voxelwise regression framework. For the VNet model, we adopt the mean squared error (MSE) loss defined as
\begin{equation}
    \mathcal{L}_{\mathrm{MSE}} = \frac{1}{DHW}\sum_i \left( Y_i^{\mathrm{pred}} - Y_i^{\mathrm{true}} \right)^2
\end{equation}

\noindent which penalises voxelwise deviations between the predicted and target DM density fields.

For the ViT-VNet model, we employ a multiscale MSE loss. In addition to the final full-resolution prediction, intermediate decoder outputs at different spatial resolutions are supervised. The total loss is expressed as

\begin{equation}
    \mathcal{L}_{\mathrm{Multi}} = \mathcal{L}_{\mathrm{Main}} + \alpha\sum_s \lambda_s \mathcal{L}_{\mathrm{MSE}}^s
\end{equation}

\noindent where $\mathcal{L}_{\mathrm{Main}}$ denotes the full resolution MSE loss, and $\mathcal{L}_{\mathrm{MSE}}^s$ represents the MSE computed at intermediate scale $s$ during convolutional down/upsampling. $\lambda_s$ are scale dependent weights, and $\alpha$ controls the overall contribution of multiscale supervision. We set $\alpha=0.2$ and $\lambda_s=2^{-s}$ for the three intermediate decoder outputs, indexed from coarse to fine as $s=1,2,3$. Thus, $(\lambda_1,\lambda_2,\lambda_3)=(0.5,0.25,0.125)$, with larger weights assigned to coarser outputs. Each intermediate prediction is interpolated to the full-resolution grid before evaluating its MSE against the target. Validation uses the final-output MSE alone.

The batch size is $24$ for both architectures. The initial learning rates are $5\times10^{-4}$ for VNet and $3\times10^{-4}$ for ViT-VNet. Each velocity input is divided by $1000\,\mathrm{km\,s^{-1}}$ before entering the network. Optimisation is performed using the Adam optimiser. The learning rate is adaptively adjusted using a $\mathrm{ReduceLROnPlateau}$ scheduler, which lowers the learning rate by half when the validation loss ceases to improve. All models are trained under identical hardware conditions, and trained $10$ times with randomised initialisation to characterise sensitivity to random initialisation.

\section{Data} \label{sec:data}

We utilise stellar kinematics combined with positional information to predict the local DM density map. Gaia provides a large number of astrometric measurements for stars in the MW, making it suitable for constructing such phase-space maps. We transform the observed astrometric measurements, consisting of positions, proper motions, and RVs, into three-dimensional stellar velocity fields. Although velocities do not directly encode the instantaneous gravitational force, they retain dynamical information about the integrated gravitational potential along stellar orbits. Interpreting this information in terms of the local DM density requires additional assumptions, such as a halo profile or a simulation-based prior. In this work, we expect the DL model to learn this mapping implicitly from the FIRE-2 training simulations.

\begin{figure}
    \centering
    \includegraphics[width=\columnwidth]{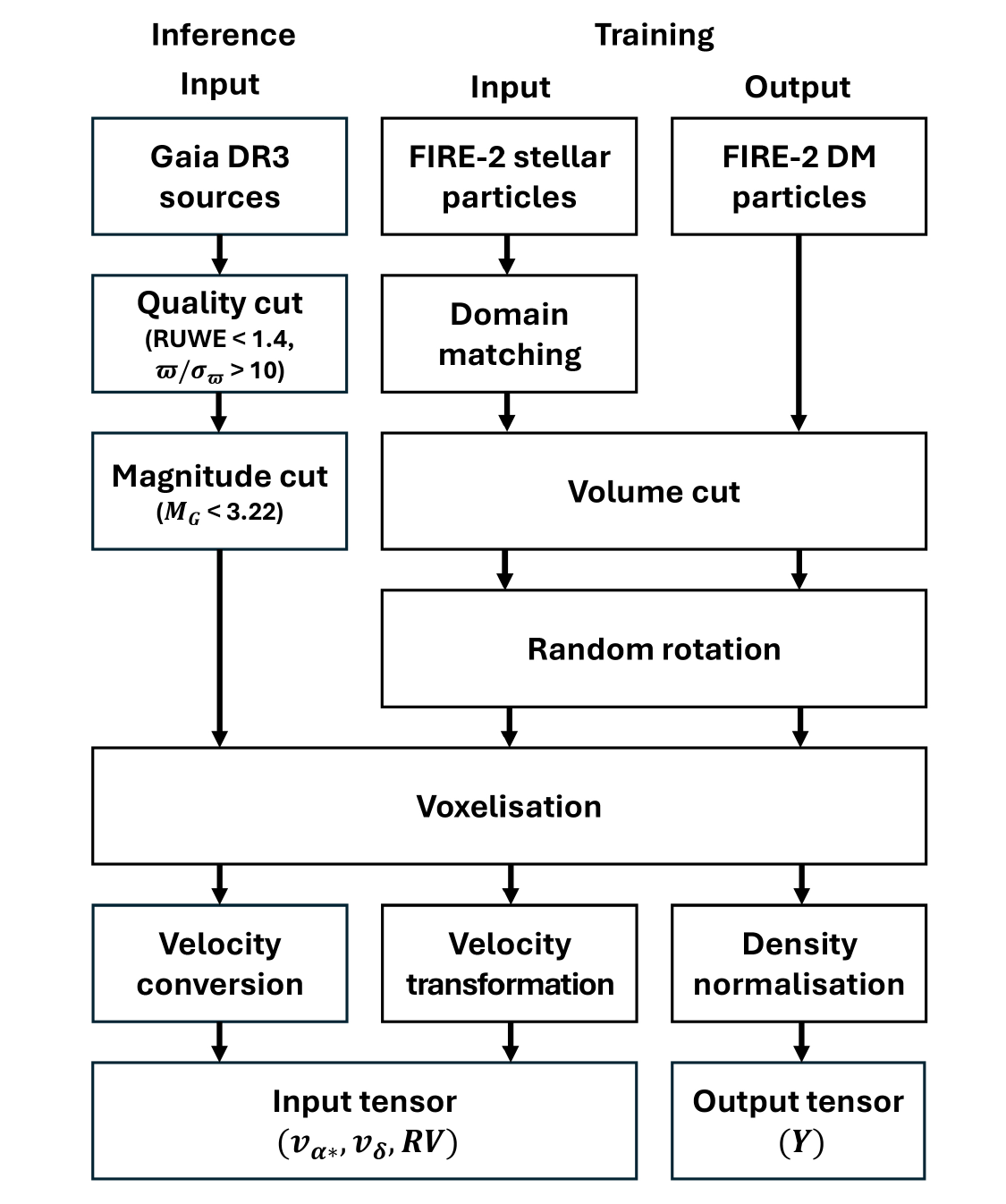}
    \caption{Schematic diagram of the data preparation pipeline used for the two architectures described in Secs.~\ref{sec:DL_vnet} and \ref{sec:DL_vit}. The right two branches illustrate the construction of the training input and output tensors from the FIRE-2 simulations, while the leftmost branch shows the construction of the inference input tensor from Gaia DR3 sources. Detailed descriptions of the FIRE-2 and Gaia data processing steps are provided in Secs.~\ref{sec:data_sim} and \ref{sec:data_gaia}, respectively. The output tensor $Y$, corresponding to the normalised DM density field, is defined in Eq.~\ref{eq:y}.}
    \label{fig:Dataprep}
\end{figure}

\subsection{FIRE-2 simulation} \label{sec:data_sim}

A large number of training samples is required to reliably train the proposed deep learning architectures. For this purpose, we adopt the FIRE-2 numerical simulation suite. FIRE-2 provides high-resolution hydrodynamic simulations of galaxies performed with the \textsc{GIZMO} code \citep{Hopkins2015, Hopkins2018_FIRE2}, making it well suited for constructing detailed phase-space training data. Its combination of high mass resolution, fine spatial resolution, and multiple MW-mass systems is particularly advantageous for our purpose, as it allows us to construct a sufficiently large set of local training volumes while limiting spatial redundancy.

Our application places stringent requirements on both mass and spatial resolution. Since the input fields are constructed from voxel-averaged stellar kinematics, a sufficiently large number of stellar particles is needed in each local volume to obtain stable phase-space estimates. At the same time, the target DM density field must be sampled by enough DM particles to avoid excessive discreteness noise before smoothing. If the particle masses were too coarse, the stellar velocity field would be sparsely sampled and the DM density target would require stronger smoothing, thereby limiting the spatial resolution of the reconstructed density field. Accurate learning of the relation between stellar kinematics and the underlying DM density therefore requires both stellar and DM particles to be resolved at sufficiently high mass resolution.

We select $7$ MW-mass galaxies from the FIRE-2 suite, labelled m12b, m12c, m12f, m12i, m12m, m12r, and m12w. Among them, m12f, m12i, and m12m are commonly referred to as the ``Latte'' simulations, as they closely resemble the MW in their global properties. In addition to these three Latte systems, we include four galaxies of comparable halo mass that have the same mass and spatial resolution, in order to broaden the range of dynamical environments represented in the training set while maintaining a uniform numerical resolution. This is important because a single simulated galaxy would provide only a limited number of approximately independent local volumes once strong spatial overlap is avoided. The use of multiple MW-mass systems therefore increases the diversity of halo structure, disk kinematics, and local dynamical conditions while maintaining the high resolution required for voxel-level training. All selected simulations share identical mass resolutions of $7.1 \times 10^{3}\,\mathrm{M}_{\odot}$ for stellar particles and $3.5 \times 10^{4}\,\mathrm{M}_{\odot}$ for DM particles. The basic physical properties of the selected galaxies are summarised in Table~\ref{tab:fire2}.

\begin{table}
\centering
\begin{tabular}{cccc}
\toprule
 & $R_{200\mathrm{m}}$(kpc) & $\mathrm{M}_{200\mathrm{m}}$ ($10^{10}\, \mathrm{M}_\odot$) & $\mathrm{M}_{*}$ ($10^{10}\, \mathrm{M}_\odot$)\\
\midrule
\midrule
MW   & $320\pm29$ & $110\pm30$ & $5\pm1$ \\
\midrule
m12b & $358$ & $143$ & $8.5$ \\
m12c & $351$ & $135$ & $5.8$ \\
m12f & $380$ & $171$ & $7.9$ \\
m12i & $336$ & $118$ & $6.3$ \\
m12m & $371$ & $158$ & $11$ \\
m12r & $321$ & $110$ & $1.7$ \\
m12w & $319$ & $108$ & $5.7$ \\
\bottomrule
\end{tabular}
\caption{Properties of the MW \citep{BlandHawthorn2016} and $7$ selected MW-like galaxies in the FIRE-2 simulation suite \citep{Wetzel2023_FIRE2}. $R_{200\mathrm{m}}$ denotes the radius enclosing a mean density $200$ times the mean matter density of the Universe, and $\mathrm{M}_{200\mathrm{m}}$ is the total mass enclosed within this radius. $\mathrm{M}_{*}$ represents the total stellar mass of each galaxy.}
\label{tab:fire2}
\end{table}

\begin{figure}
    \centering
    \includegraphics[width=\columnwidth]{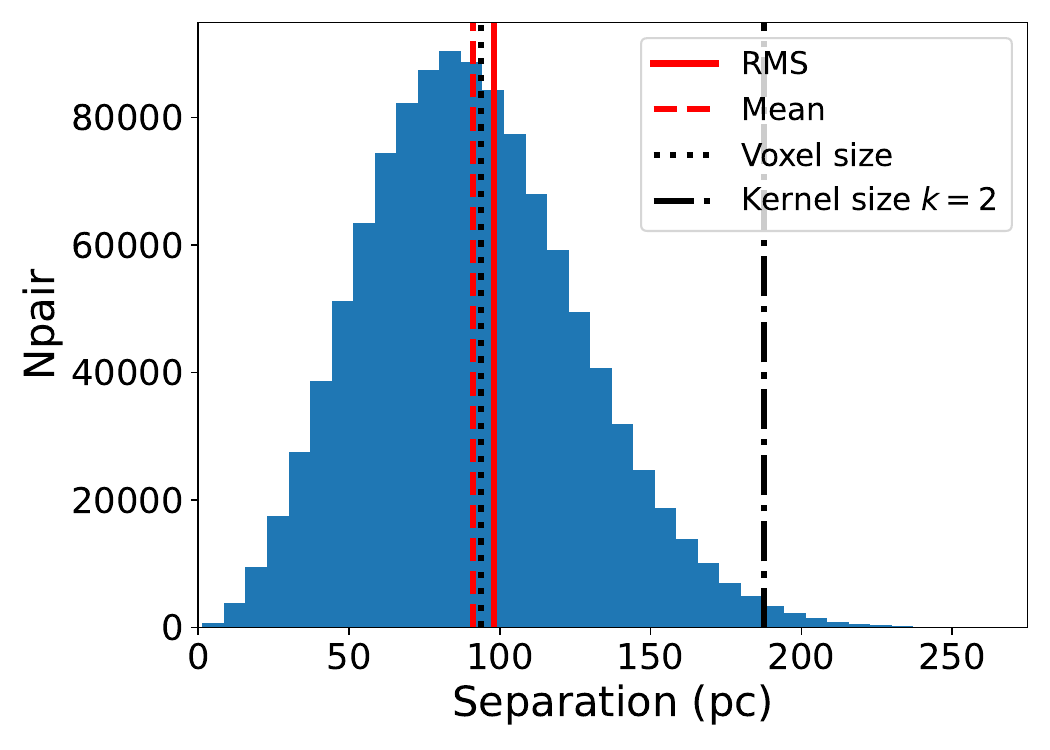}
    \caption{Histogram of particle-to-particle separation for DM particle pairs in the FIRE-2 simulated galaxies used in this study. The DM particles were chosen in the disc region with $R = 8.2 \pm 3 \, \mathrm{kpc}$. The black dash-dotted line corresponds to the Gaussian standard deviation, $\sigma=k=2$ voxels, and the dotted line corresponds to the length of the voxel edge. The red solid and dashed vertical lines indicate the root-mean-square (RMS) and mean of particle separation, respectively.}
    \label{fig:dmsep}
\end{figure}

Given the volumetric architectures described in Sec.~\ref{sec:DL_vnet} and \ref{sec:DL_vit}, both stellar and DM particles in the simulation must be transformed into a structured three-dimensional representation. To construct the output, we discretise the $(6\,\mathrm{kpc})^3$ volume into a $64^3$ voxel grid, resulting in a spatial resolution of $(93.75\,\mathrm{pc})^3$ per voxel. DM particles are assigned to voxels and converted into discrete mass density by dividing the summed mass of DM particles in each voxel by the voxel volume. To obtain an output tensor suitable for regression, the discrete density map is first smoothed using a Gaussian kernel with standard deviation $\sigma=k$ (in voxel units). We adopt $k=2$, corresponding to $\sigma=187.5\,\mathrm{pc}$ (FWHM $\simeq442\,\mathrm{pc}$) and safely exceeding both the gravitational softening length of the DM particles in FIRE-2 simulations ($\epsilon_{\mathrm{DM}}=40\,\mathrm{pc}$, fixed in physical units at the epochs considered; \citealt{Hopkins2018_FIRE2, Wetzel2023_FIRE2}) and the typical particle-to-particle separation of DM particles in the disc region of $R = 8.2 \pm 3 \, \mathrm{kpc}$ ($\sim 100 \, \mathrm{pc}$, see Fig.~\ref{fig:dmsep}), avoiding excessive smoothing of small-scale density structures.

\begin{equation}
    Y = 
    \frac{\log_{10}\left( \rho_{\mathrm{DM}}/{1\,M_\odot\,{\rm pc}^{-3}} + 1 \right)}
         {\log_{10}\left( \rho_{\mathrm{DM,max}}/{1\,M_\odot\,{\rm pc}^{-3}} + 1 \right)}.
    \label{eq:y}
\end{equation}

We then normalise the DM density field using a monotonic log1p-based transformation that maps the target values onto a bounded interval. Since the DM densities in our training set are generally below the reference density of $1\,M_\odot\,{\rm pc}^{-3}$ and span a modest dynamic range, the mapping is close to linear over most of the relevant density range, while remaining well defined at $\rho_{\rm DM}=0$. Here, $\rho_{\rm DM,max}$ is the maximum DM density computed across the training, validation and test sets, $0.346\,\mathrm{M}_{\odot}\,\mathrm{pc}^{-3}\simeq13.1\,\mathrm{GeV\,cm^{-3}}$. The division by the reference density makes the argument of the logarithm dimensionless, and the additive constant ensures that zero-density voxels are mapped to a finite value. The resulting target field lies within the limited range of $Y \in [0,1]$, to ensure a stable training procedure. The physical density field is recovered by inverting the above transformation: 
\begin{equation}
    \rho_{\mathrm{DM}} 
    = \left[ 
    10^{\, Y \times \log_{10}\left(\rho_{\mathrm{DM,max}}/{1\,M_\odot\,{\rm pc}^{-3}} + 1\right)} 
    - 1 
    \right]\,M_\odot\,{\rm pc}^{-3}.
\end{equation}

The input stellar velocity field is constructed on the same voxel grid as the output. Stellar particles are first assigned to voxels according to their positions. Within each voxel, we compute the mean stellar velocity, producing a three-dimensional kinematic field that serves as the input tensor to the network. To construct localised training samples comparable to the solar neighbourhood, we extract voxelised cubic volumes across the galactic disc plane while converting the Cartesian velocities of stellar particles into tangential/radial velocities. For each sample, the reference stellar particle nearest the volume centre defines the simulated observer. The reference particle's position and velocity are subtracted from those of the surrounding particles before the resulting relative velocities are projected into observer-centred tangential and radial components. Thus, the observer moves with the reference particle rather than remaining at rest in the Galactocentric frame. We exclude the innermost $2.2\,\mathrm{kpc}$ region near the bulge to avoid strong kinematic deviations from disc-dominated environments. Under this configuration, $24$ volumes can be extracted from a single galaxy while minimising the volume overlapping.

To further enlarge the dataset, we extend the sample along the time dimension. For each galaxy, we include $11$ snapshots separated by approximately $22\,\mathrm{Myr}$, including the present epoch. We confirmed that both the stellar and DM particles occupying a given volume are substantially replaced over this interval. Nevertheless, we acknowledge that the gravitational potential and the resulting stellar motions may remain temporally correlated. In addition, each cube is subjected to random three-dimensional rotations about the simulation's Cartesian $x$, $y$, $z-$axes, generating four additional orientations alongside its original configuration. This augmentation preserves the intrinsic dynamical structure while increasing the diversity of spatial configurations. Combining all procedures yields a total of $9240\,\,(= 7\times24\times11\times5)$ volumetric samples derived from $168\,\,(= 7\times24)$ spatially distinct volumes across the seven systems.

The simulated galaxy `m12m', including all of its snapshots, spatial volumes, and rotated realisations, is reserved exclusively for testing. Leaving a single galaxy to test the trained model ensures that the test set shares neither spatially overlapping volumes nor temporally correlated snapshots with the training and validation sets, providing a robust assessment of generalisation to an unseen galaxy. The remaining six galaxies were used to generate the training and validation sets with random division, by a ratio of $8:2$. This results in approximate training, validation, and test fractions of $69:17:14$ per cent, respectively. During random division between the training and validation sets, different snapshots and rotations of the same volume may occur in both sets. However, all m12m realisations are held out for testing, and the consistent test results suggest that this overlap does not inflate performance on the unseen galaxy. The pipeline for generating each cube is summarised and schematised in the second and third columns of Fig.~\ref{fig:Dataprep}.

Generating synthetic stellar catalogues that mimic Gaia observations is possible through the Ananke framework built upon FIRE-2 \citep{Sanderson2020, Nguyen2024_Ananke}. This framework distributes synthetic stars according to the underlying phase-space structure of the simulated galaxy. However, in adequately sampled voxels, mock sampling is expected to preserve much of the parent mean velocity field, and our methodology relies on voxel-averaged kinematic fields; we thus directly utilise the original stellar particles from the simulation suite. We note that selection, extinction and observational errors can nevertheless alter that field and its coverage.

\subsection{Gaia DR3 nearby star catalogue} \label{sec:data_gaia}

\begin{figure}
    \centering
    \includegraphics[width=\columnwidth]{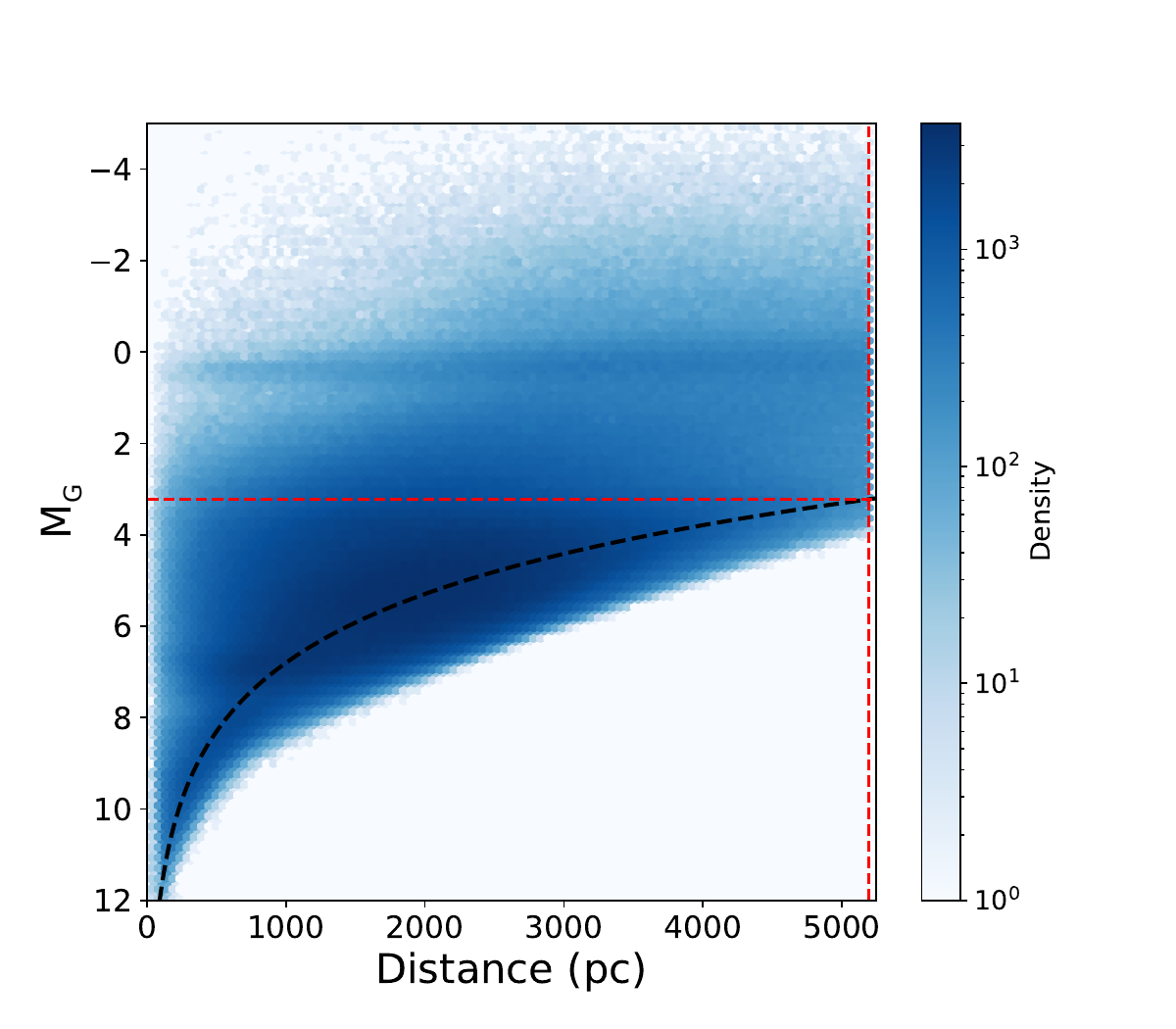}
    \caption{Distances versus absolute magnitude 2D histogram of Gaia DR3 stars. The cutoff in the lower part of the plot corresponds to the magnitude limit. The black dashed curve is the nominal apparent-magnitude boundary (approximately $G=16.8$ without extinction). Red dashed lines indicate the adopted luminosity and distance limits.}
    \label{fig:gaia_dist_mag}
\end{figure}

\begin{figure*}
    \centering
    \includegraphics[width=\textwidth]{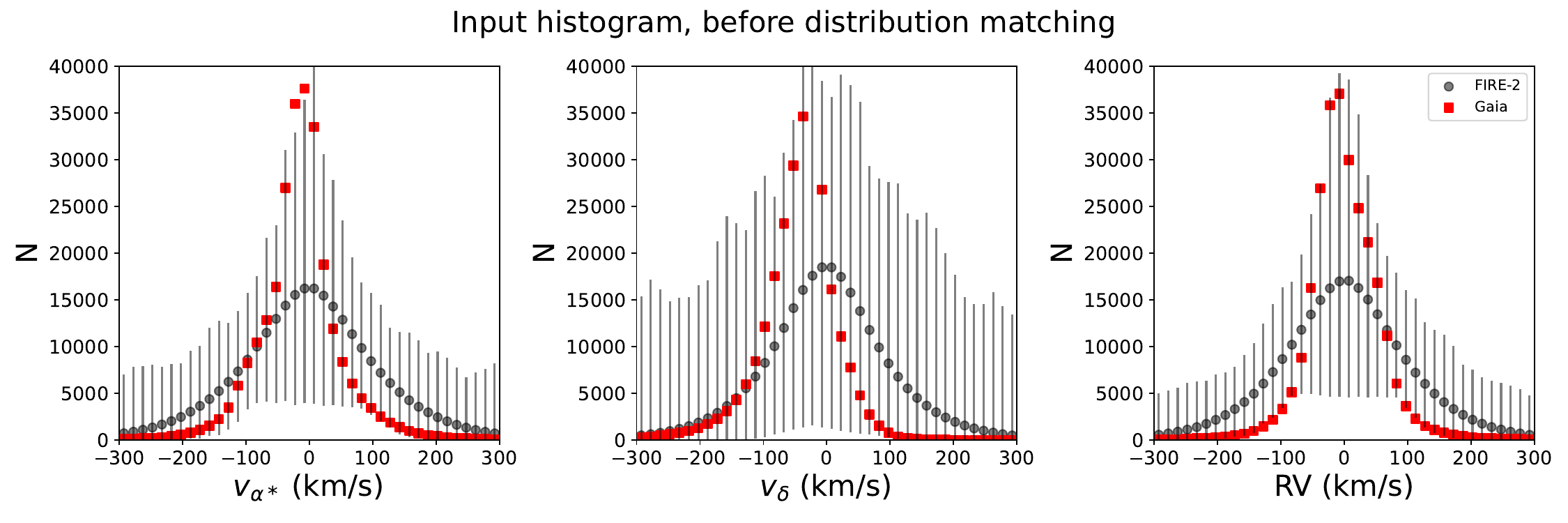}
    \includegraphics[width=\textwidth]{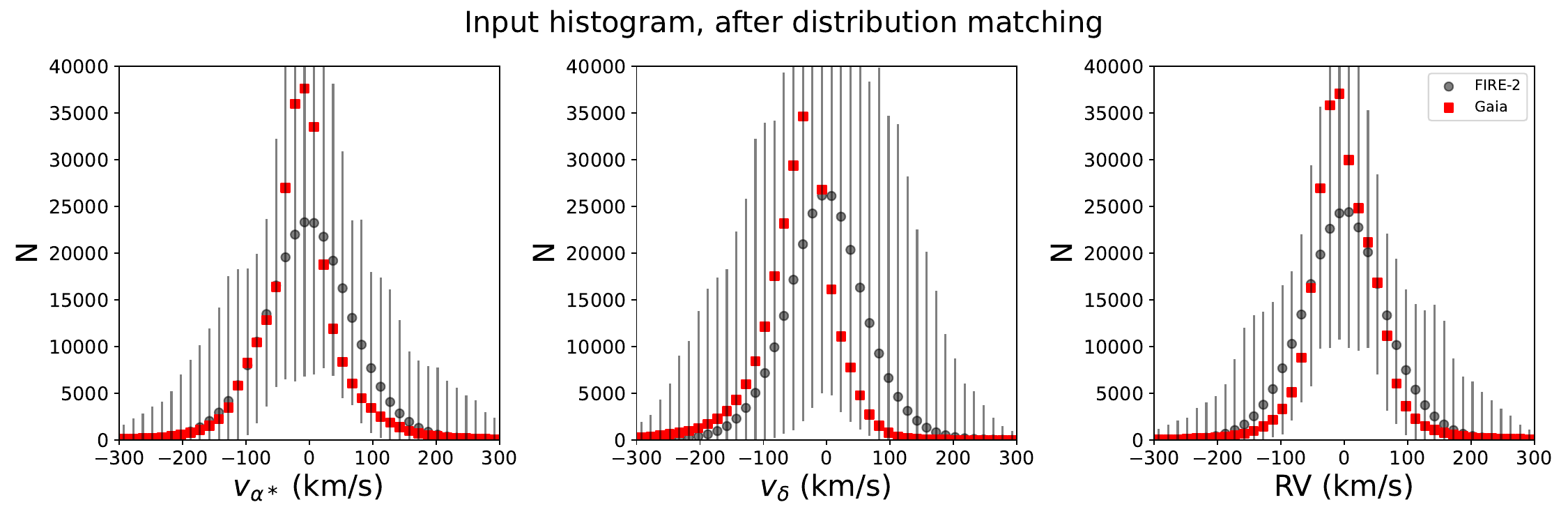}
    \caption{Histograms of the input channels: tangential velocities along the R.A. and Decl. directions ($v_{\alpha*}$ and $v_{\delta}$), and RV. Red squares show the corresponding distributions constructed from the Gaia sample after applying the selection criteria described in Sec.~\ref{sec:data_gaia}. Grey circles with error bars in the top panels represent the distribution of all input values $X \in \mathbf{X}$ in the training set before applying the Gaia-matching cuts. For each bin, the error bars span the minimum and maximum counts over the training volumes, and therefore indicate the full range covered by the simulated training set rather than statistical uncertainties. They are shown to illustrate whether the Gaia distributions fall within the support of the training set. Bottom panels show the same comparison after applying the domain matching, using the Gaia-matched subsample.}
    \label{fig:inputs_hist}
\end{figure*}

We use the Gaia DR3 release to obtain astrometric measurements of nearby stars. The full source catalogue contains more than $1.8$ billion objects, among which $585$ million have five-parameter astrometric solutions, comprising right ascension ($\alpha$), declination ($\delta$), parallax ($\varpi$), and proper motions ($\mu_{\alpha*}$ and $\mu_{\delta}$). Six-parameter astrometric solutions, which additionally include pseudo-colour, are available for $882$ million objects. Note that the proper motion along the R.A. direction includes a factor of $\mathrm{cos}\delta$, as $\mu_{\alpha*} = \mu_{\alpha}\mathrm{cos}\delta$. RVs are available for approximately $34$ million objects \citep{GaiaDR3_summary}.

From the catalogue including both five- and six-parameter astrometric solutions, we select sources suitable for constructing the input voxelised phase-space maps. We apply $\mathrm{RUWE}<1.4$ and $\varpi/\sigma_{\varpi}>10$. Photogeometric distances from \citet{BailerJones2021} are used where available; otherwise we invert parallaxes corrected using \citet{Lindegren2021}.

To construct a sample with reduced distance-dependent photometric incompleteness within a cubic region of side length $6\,\mathrm{kpc}$ centred on the Sun, we impose an absolute magnitude limit. We use the Gaia G-band magnitude corrected for interstellar extinction using the available GSP-Phot extinction estimates provided in the catalogue. Considering that the farthest point within the cube lies at a distance of $3\sqrt{3}\,\mathrm{kpc}$, we adopt a luminosity threshold of $M_G<3.22$ (see Fig.~\ref{fig:gaia_dist_mag}).

For the selected sources, we convert the angular proper motions into physical tangential velocity components along the R.A. and Decl. directions using the adopted distances. We denote these components as $v_{\alpha *}$ and $v_{\delta}$, respectively. The input features are therefore constructed from $(v_{\alpha *}, v_{\delta}, {\rm RV})$, consistent with the velocity components used in the training set.

After applying all quality cuts and magnitude limits, the final Gaia sample contains $13\,544\,206$ sources, including $7\,259\,188$ with RV measurements. When mapped onto the adopted voxel grid, this corresponds to an average of approximately $52$ sources per voxel for the tangential channels and $28$ for RV; actual occupancy varies with position within the Galaxy. We note that RV measurements are available for a subset of Gaia DR3 sources and have different selection functions \citep{CantatGaudin2023, CastroGinard2023}. Therefore, the RV channel was constructed using only those sources in the final selected sample with available RV measurements. The pipeline for generating an input tensor from the Gaia DR3 source catalogue is shown in the first column of Fig.~\ref{fig:Dataprep}.

\subsection{Selection effects} \label{sec:method_sel}

Naturally, selection criteria described in Sec.~\ref{sec:data_gaia} preferentially retain relatively bright stars. This introduces a potential bias in the spatial and kinematic distribution of the sample across the cubic volume. The luminosity cut selects both bright young stars and evolved red-clump and red-giant stars, and therefore does not necessarily favour a particular component of the MW. Nevertheless, it may alter the mixture of dynamical populations, so the selected sample should not be assumed to represent the full stellar population across the volume. As a result, the selected Gaia sample may not uniformly trace all dynamical components within the volume.

Additional selection cuts, such as colour-based filtering, could partially mitigate this imbalance by targeting more homogeneous stellar populations. However, such constraints would restrict the sample to specific stellar populations, introducing additional assumptions about their dynamical representativeness. We therefore retain a broader stellar sample without additional population-specific cuts, allowing us to investigate the effects of subsequent tracer sampling on the inferred DM density.

Interstellar extinction further complicates the spatial completeness of the sample, particularly along the Galactic disc where dust attenuation is significant. Because the Gaia catalogue and associated extinction coefficients do not allow a direct reconstruction of intrinsically obscured stellar populations, it is difficult to distinguish between genuine underdensities and extinction-induced incompleteness. Although the stellar number density field contains additional dynamical information, incorporating it into the observational analysis would require strongly model-dependent completeness corrections. An ablation test confirmed that adding the stellar number-density field as an input channel gives a small but statistically resolved improvement in the VNet test performance as shown in Appendix~\ref{sec:abltest}. Nevertheless, we adopted the velocity-only configuration because the observational stellar number-density field is strongly affected by extinction and spatial incompleteness, which would introduce additional model-dependent systematics when applying the network to Gaia data.

The selected Gaia sample exhibits narrower velocity distributions than the unselected FIRE-2 stellar population. This can reflect both observational selection and intrinsic dynamical differences, rather than the luminosity cut alone. The resulting velocity distributions are shown in the top panels of Fig.~\ref{fig:inputs_hist}, which highlight the discrepancy between the Gaia sample and the raw stellar distributions in the FIRE-2 simulations.

To mitigate potential out-of-distribution effects arising from this discrepancy, we apply a kinematic rejection-sampling procedure to the FIRE-2 stellar particles. We first transform the stellar velocities in both the Gaia sample and the FIRE-2 simulations into a common Galactocentric velocity space and use the cylindrical components, $(v_R, v_\phi, v_z)$, for the matching. In this transformation, we adopt the total Cartesian Galactocentric solar velocity $\boldsymbol{v}_{\odot,\mathrm{GC}}=(12.9,245.6,7.78)\,\mathrm{km\,s^{-1}}$, with $R=8.122\,\mathrm{kpc}$, using \texttt{astropy} \citep{Astropy2022}. From the selected Gaia sample, we construct one-dimensional reference histograms for each velocity component. The same histograms are constructed for the FIRE-2 stellar particles, and each simulated stellar particle is assigned an acceptance probability $P_{\rm acc}$. For each simulated galaxy, this probability is evaluated galaxy-wide immediately after loading the stellar particles, before extracting spatial cubes and voxelising the inputs.
\begin{equation}
    P_{\rm acc} =
    \min\left[
    1,
    \prod_{i \in {R,\phi,z}}
    \frac{p_{\rm Gaia}(v_i)}{p_{\rm FIRE}(v_i)}
    \right]
    \label{eq:pacc}
    \end{equation}
\noindent After rejecting dynamically hotter stellar particles according to $P_{\rm acc}$, the matched FIRE-2 subsample is then used to construct the stellar input fields, while the corresponding DM density fields are computed from the original DM particles in the same spatial volumes. The bottom panels of Fig.~\ref{fig:inputs_hist} show the resulting matched distributions in the input velocity space. The effects of this domain-matching procedure are discussed in Appendix~\ref{app:kinematics}.

We additionally downsample the Gaia sources to bring the tracer sampling of the observational inputs closer to that of the simulated inputs. Because the voxelwise velocities are averages over finite numbers of tracers, differences in tracer counts affect the sampling fluctuations and the occurrence of empty voxels, even when the underlying velocity distributions are similar. We therefore use the vertical sampling pattern, which is obtained by counting the fraction of empty voxels in each slice along the direction perpendicular to the galactic plane in the m12m test samples, as a reference for randomly downsampling the Gaia sources before constructing the input velocity fields. The sampling prescription and its effect on the reconstructed density are described in Appendix~\ref{app:counts}.

\section{Results} \label{sec:result}

\subsection{Machine performance test} \label{sec:training}

\begin{figure}
    \centering
    \includegraphics[width=\columnwidth]{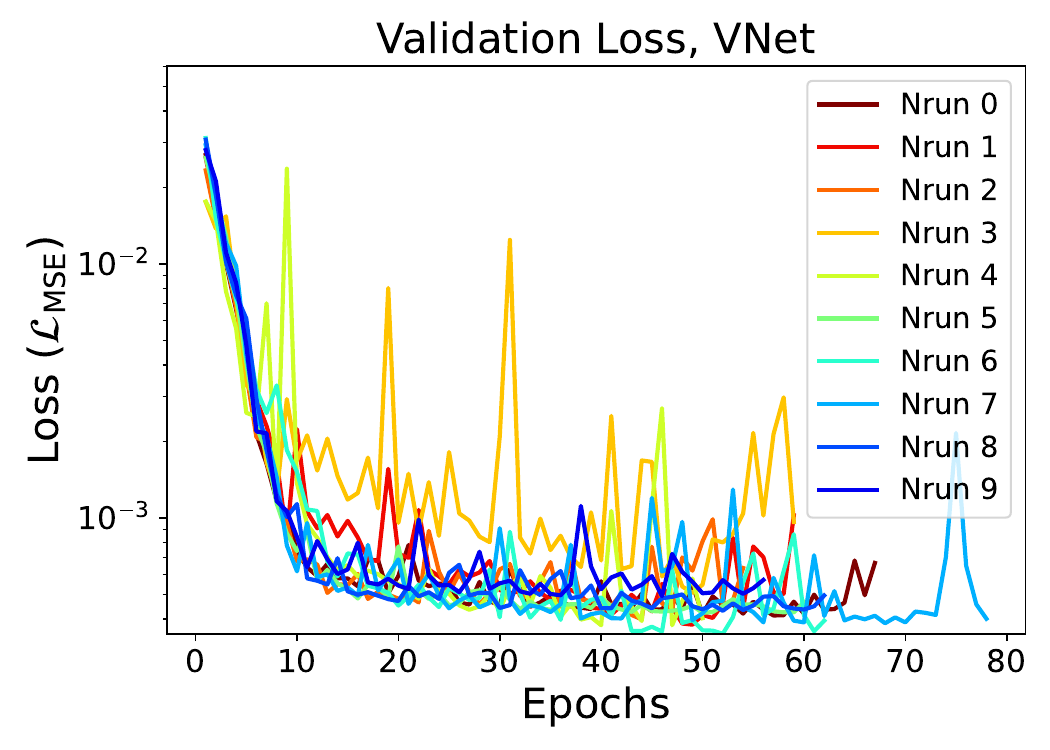}
    \includegraphics[width=\columnwidth]{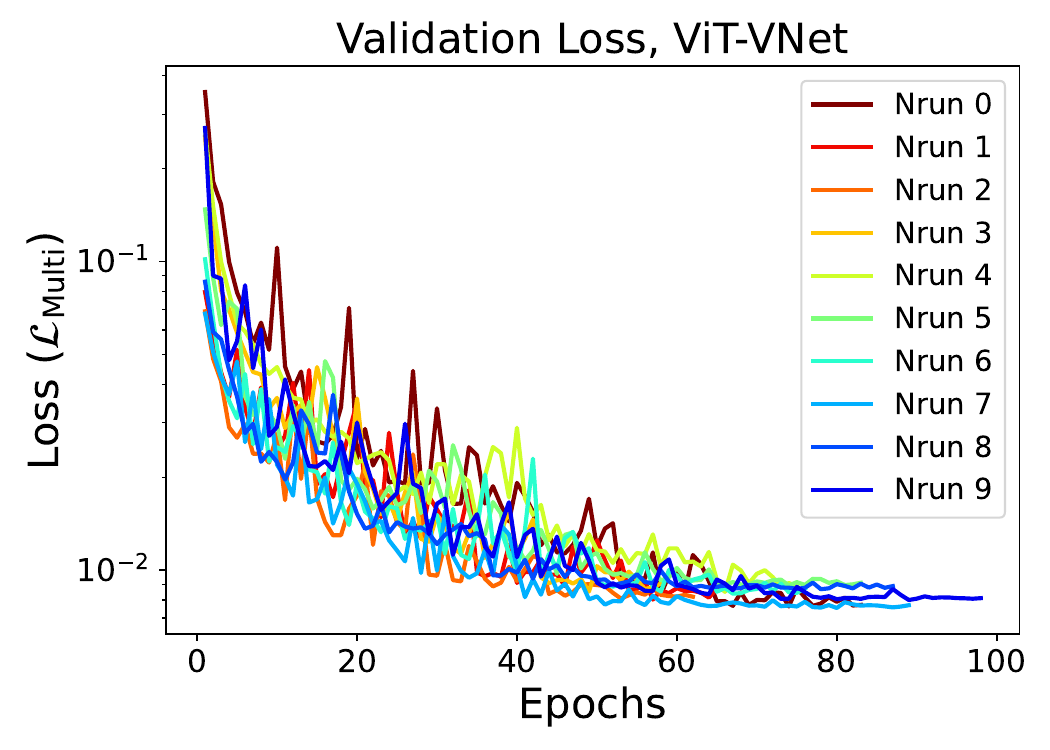}
    \caption{Validation loss curves for the two architectures employed in this study, VNet and ViT-VNet. Colours denote $10$ independent training runs per architecture. Training was stopped when the validation loss failed to improve following learning rate reduction.}
    \label{fig:valloss}
\end{figure}

\begin{figure*}
    \centering
    \includegraphics[width=170mm]{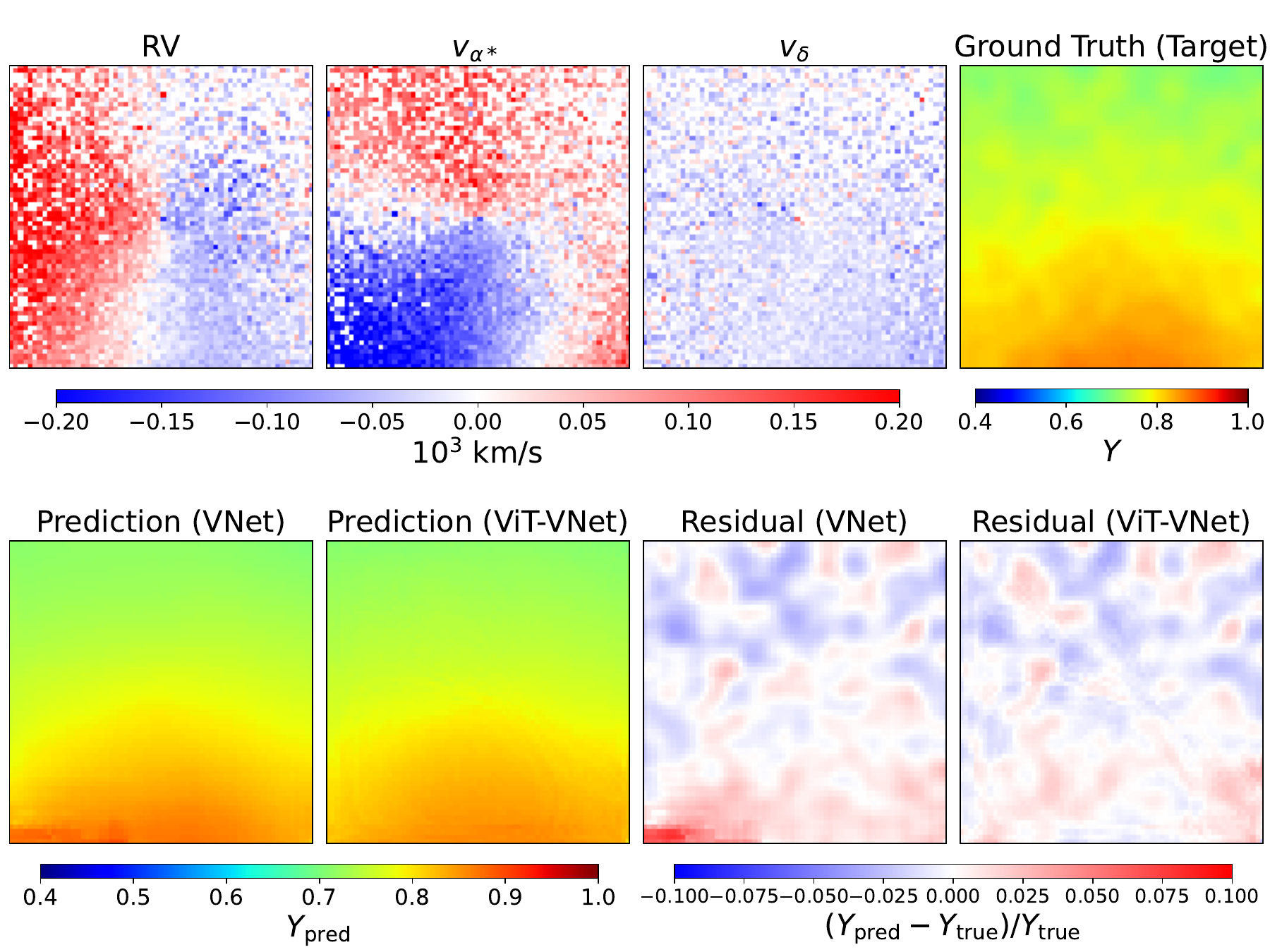}
    \caption{Input features and predicted DM density fields for a randomly selected test sample. Each panel spans $6\,\mathrm{kpc}$ on a side, with the simulated galactic centre located towards the bottom of each panel. The first row shows a single depth slice of the input features and the ground-truth DM density field. The left two panels in the second row present the mean prediction obtained from $10$ independently trained models for each architecture. The right two panels in the second row display the fractional residual, defined as $(Y_{\mathrm{pred}} - Y_{\mathrm{true}})/Y_{\mathrm{true}}$, highlighting spatial deviations between the prediction and the simulation.}
    \label{fig:train_vis}
\end{figure*}

We first assess the prediction fidelity of the two architectures by training each architecture $10$ times with different random initialisations. In each training run, we reduced the learning rate by half when the validation loss did not decrease by more than 0.01 per cent relative to the lowest loss over the previous 5 consecutive epochs. If the validation loss did not improve for 10 epochs corresponding to two learning rate reduction steps, we stopped the training loop. The resulting loss convergence for each training run is shown in Fig.~\ref{fig:valloss}. We selected the model marking the lowest validation loss for each training loop. 

This procedure yields $10$ predictions for each test sample, providing 10 predicted values at each voxel position. For both the test samples and the Gaia sample, we adopt the arithmetic mean of the ten predictions at each voxel as the representative predicted density, separately for each architecture. Fig.~\ref{fig:train_vis} presents a representative one-voxel-thick slice from the test set. The slice is located at the centre of the volume, sliced perpendicular to $z$-axis. Visually, the predictions from VNet and ViT-VNet are nearly indistinguishable, both reproducing the density gradient over the whole volume. This suggests that both architectures converge to a similar mapping between stellar phase-space information and the underlying DM density distribution. However, both architectures fail to reproduce the small-scale structures.

\begin{figure}
    \centering
    \includegraphics[width=80mm]{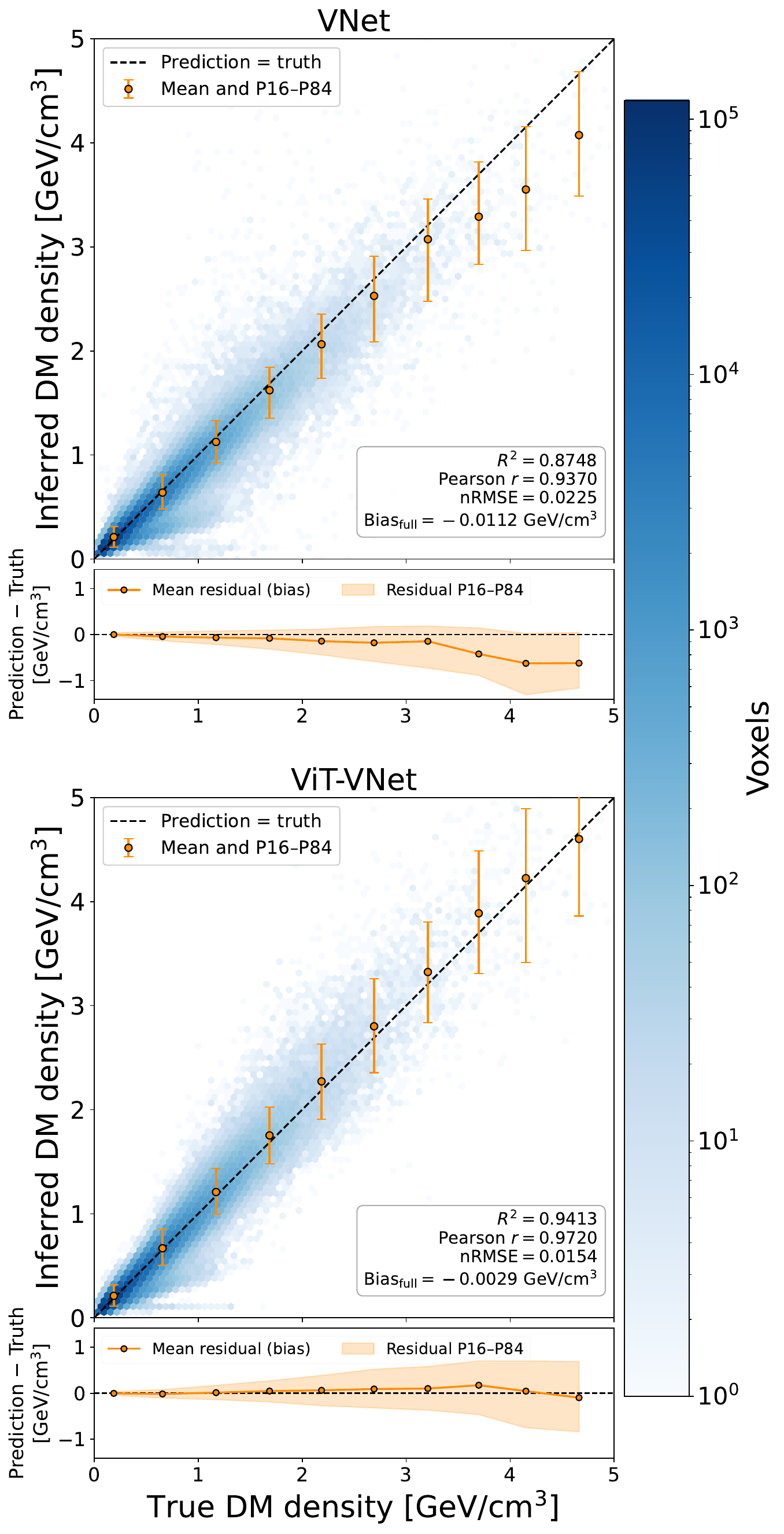}
    \caption{Voxel-to-voxel value comparison plot to show the relation between true value and predicted value of DM density. All values are expressed in physical DM density units. Orange markers and intervals show the means and 16th--84th percentiles of predictions within bins of true density, combining ten realisations per architecture. The residual panel below each comparison shows the mean signed residual, $\rho_{\rm pred}-\rho_{\rm true}$, with the shaded band indicating its 16th--84th percentile range. Statistical performance metrics for each architecture are marked on corresponding panels. The coefficient of determination ($R^2$) and the Pearson correlation coefficient ($\mathrm{Pearson}\,\, r$) quantify the strength of the linear agreement between the predictions and the ground truth, with values closer to unity indicating better predictive performance. The normalised RMSE is $\mathrm{RMSE}/[\max(\rho_{\rm true})-\min(\rho_{\rm true})]$, where lower values correspond to smaller prediction errors. The full-sample bias, $\mathrm{Bias}_{\rm full}$, is the mean signed residual over the full test set, with negative values indicating underprediction.}
    \label{fig:train_v2v}
\end{figure}

To quantify the predictive performance, we directly compare the voxelwise predicted values with the ground truth (See Fig.~\ref{fig:train_v2v}). Both architectures closely follow the one-to-one relation, while the binned means and residual panels show density-dependent bias and greater scatter at higher densities. The coefficient of determination, the Pearson correlation coefficient and normalised RMSE, are also summarised in the same figure. We characterise density-dependent prediction deviations using the 16th--84th percentile range of the held-out m12m test predictions, combining the outputs of all ten models within bins of true density. We use this range as a test-based estimate of prediction scatter at the density inferred from the Gaia sample.

\subsection{3D DM density map}  \label{sec:map}

\begin{figure*}
    \centering
    \includegraphics[width=\textwidth]{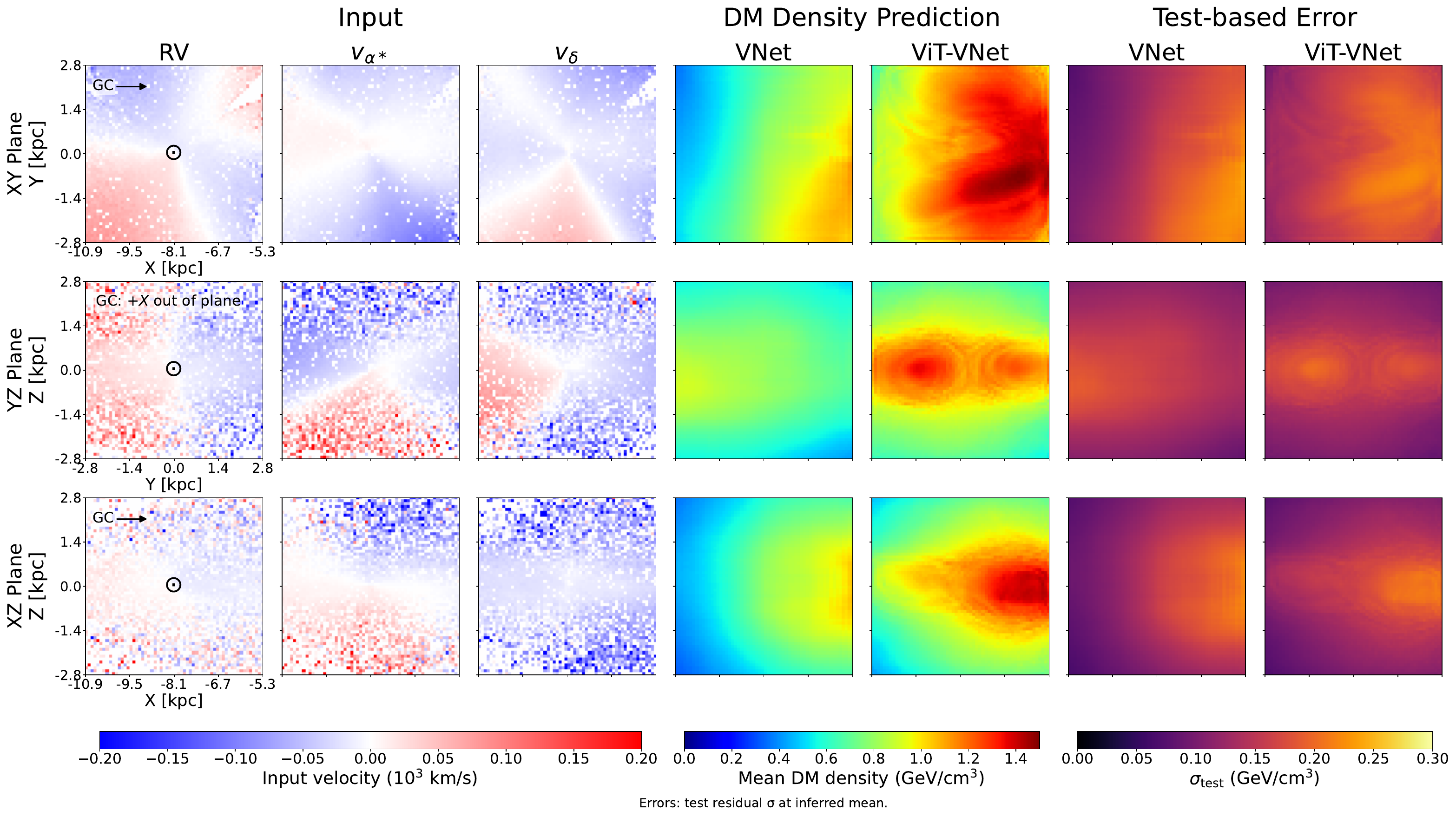}
    \caption{Predicted DM density field and test-based prediction-deviation diagnostics in the solar neighbourhood, reconstructed using VNet (fourth column) and ViT-VNet (fifth column). Each row presents a planar slice in the galactocentric coordinate system, corresponding to the $XY$, $YZ$, and $XZ$ planes (from top to bottom). The Sun is located at the centre of each panel, and the direction towards the Galactic centre is indicated by a solar symbol ($\odot$) and an arrow in the panels in the first column. The sixth and seventh columns show test-based prediction-deviation diagnostics for VNet and ViT-VNet, respectively; these are evaluated at the inferred ensemble-mean density and are not complete observational uncertainties. The input features are displayed in the first three columns. The predictions use the arithmetic mean of ten models per architecture, with voxel downsampling.}
    \label{fig:gaia_pred}
\end{figure*}

We then applied the Gaia sample to the trained models to obtain the predicted DM density field in the solar neighbourhood. The resulting volumetric density maps are shown in Fig.~\ref{fig:gaia_pred}. Each subplot presents a one-voxel-thick slice of the input features and the reconstructed density field in the Galactocentric coordinate system, along the $XY$, $YZ$, and $XZ$ planes (from top to bottom).

The fourth and fifth columns display the predicted DM density in physical units of $\mathrm{GeV\,cm^{-3}}$. Although the two architectures showed nearly indistinguishable performance on the test sets in both visual and statistical evaluations, their predictions from the Gaia input exhibit noticeable structural differences. Nevertheless, both models consistently recover a global radial gradient, with the density increasing towards the Galactic centre (GC) and decreasing outward, qualitatively consistent with expectations from standard DM halo models.

The sixth and seventh columns show test-based prediction deviations evaluated at the inferred ensemble-mean density. For each architecture, these are derived from the 16th--84th percentiles of the held-out m12m predictions within bins of true density, pooling the outputs of the ten models trained with different random initialisations. The resulting ranges therefore include both variation among models and prediction scatter across the test samples. To further quantify the reconstructed DM field in a compact and directly comparable form, we next consider a scalar estimate of the local DM density in the solar neighbourhood.

\begin{figure}
    \centering
    \includegraphics[width=\columnwidth]{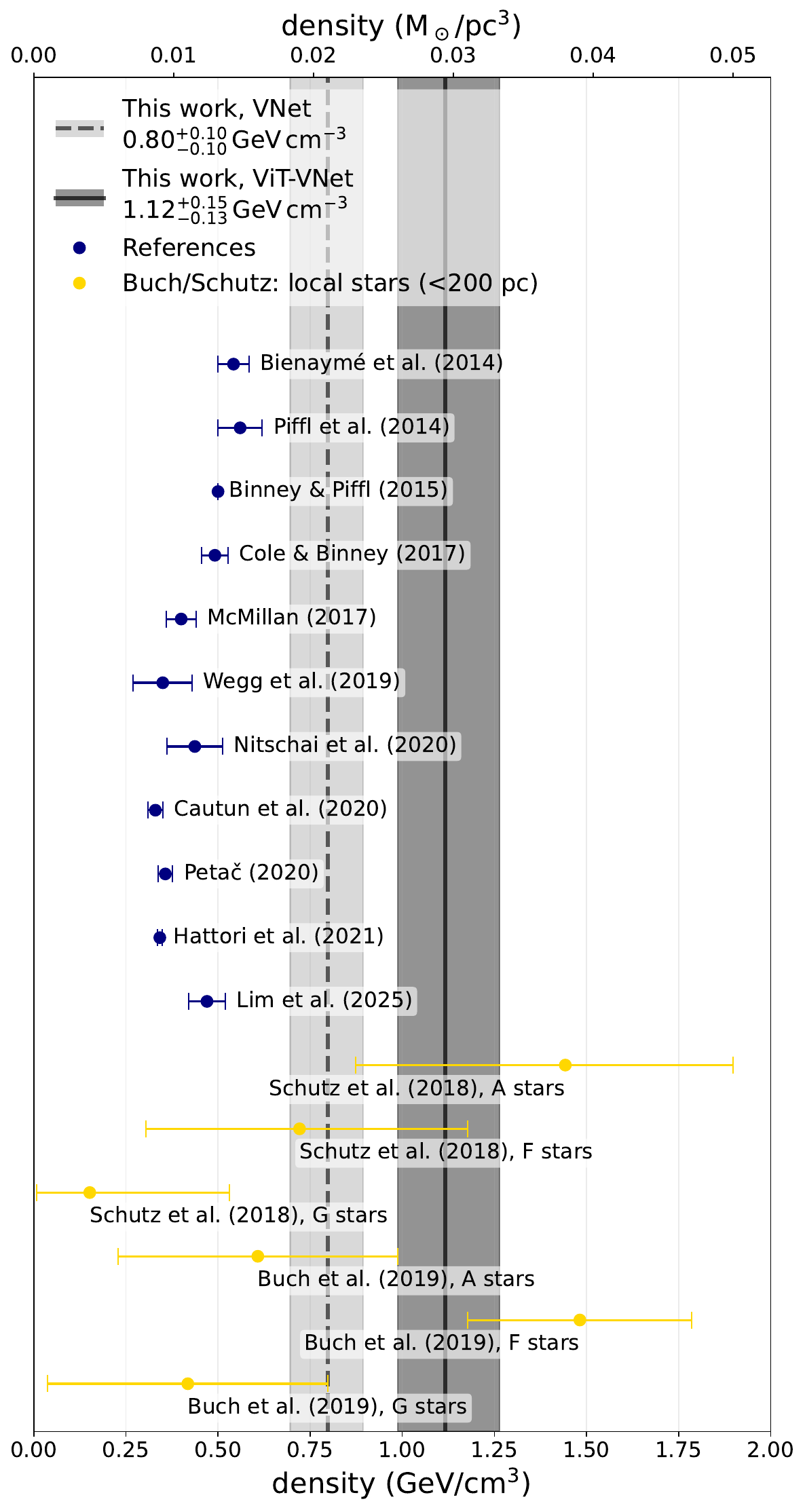}
    \caption{Comparison of the solar neighbourhood DM density between previous studies and our predictions. The vertical dashed and solid lines indicate the VNet and ViT-VNet predictions, measured within the eight central voxels representing the solar neighbourhood. Grey bands show the associated test-based prediction-deviation diagnostics. Coloured markers correspond to values reported in the literature. The yellow markers represent measurements using stars only within $200\,\mathrm{pc}$. The predicted values are shown in two units: $\mathrm{GeV}\,\mathrm{cm}^{-3}$ and $\mathrm{M}_{\odot}\,\mathrm{pc}^{-3}$.}
    \label{fig:gaia_pred_comp}
\end{figure}

\subsection{Local DM density}  \label{sec:scala}

To estimate the local DM density from the reconstructed field, we focus on the central voxels surrounding the solar position. Since the reconstruction grid has an even number of cells along each axis and the observer is placed at the centre of the volume, the solar position lies at the common corner of the $8$ central voxels. We therefore select these $8$ voxels, which form a cube with an edge length of $187.5 \, \mathrm{pc}$. For each architecture, we compute the mean density within this region from the ten-model mean field. The quoted deviations are evaluated from the held-out test predictions at the inferred density.

Our best estimation of solar neighbourhood DM density is $0.80^{+0.10}_{-0.10}$ and $1.12^{+0.15}_{-0.13}\,\mathrm{GeV}\,\mathrm{cm}^{-3}$ from VNet and ViT-VNet. The reported uncertainty ranges correspond to the inference scatter obtained from the test set (Fig.~\ref{fig:train_v2v}). The $0.32\,\mathrm{GeV}\,\mathrm{cm}^{-3}$ difference between architectures is reported separately from these test-based deviations. Sensitivities associated with tracer sampling and domain matching are also discussed separately (Appendix~\ref{app:B}), suggesting that the absolute normalisation requires further study. These values are presented in Fig.~\ref{fig:gaia_pred_comp}, and compared with previous measurements reviewed in \citet{deSalas2020}. Our DL-based estimates exceed all the non-local determinations shown in Fig.~\ref{fig:gaia_pred_comp} \citep{Bienayme2014, Piffl2014, Binney2015, Cole2017, McMillan2017, Wegg2019, Nitschai2020, Cantun2020, Petac2020, Hattori2021, Lim2025}. Local measurements based on A-, F-, and G-type stars within $\sim 200 \, \mathrm{pc}$ \citep{Schuts2018, Buch2019} are also shown in the same plot.

\section{Discussion} \label{sec:discuss}

\subsection{Small scale predictability} \label{sec:discuss1}

We first discuss the predictability of small-scale structures in the DM density field. As shown in Fig.~\ref{fig:train_vis}, small fluctuations are less prominently reconstructed than the large-scale density gradient. Because the architectures are trained to learn the mapping between stellar kinematics and the underlying DM density field, the training outcome should be interpreted in terms of how strongly small-scale density variations are encoded in stellar phase-space information.

The construction of the target density field in training set directly affects the inference. The discrete DM particle distribution is smoothed using a Gaussian kernel to obtain a continuous field suitable for regression. As the adopted kernel size of $k=2$ voxels already exceeds both the gravitational softening length and the typical separation between DM particle pairs, increasing the smoothing scale further would artificially erase physically possible small-scale density structures. Conversely, reducing it would amplify discreteness noise. Therefore, the recoverable small-scale structure is intrinsically limited by the resolution and smoothing scale of the simulation.

More fundamentally, stellar kinematics may not respond strongly to small-scale density perturbations. Stellar velocities reflect the time integrated gravitational acceleration along their orbits rather than instantaneous local forces. As a result, small-scale density fluctuations can be dynamically averaged out over orbital timescales, weakening their imprint on the observable phase-space distribution. Even if small-scale density fluctuations exist in the simulation, their kinematic signatures may therefore be intrinsically suppressed.

In addition, both the target system (the MW) and the training galaxies are late-type disc galaxies, and our volumes are extracted from disc dominated regions. In such environments, circular rotation constitutes the dominant component of stellar motion. The large rotational velocity may dominate over non-circular or locally perturbed motions, reducing the diversity of dynamical responses available for learning small-scale density fluctuations.

Taken together, these considerations suggest that the limited recovery of small-scale density structures is not solely a shortcoming of the DL architectures, but reflects a combination of simulation resolution and the physical coupling between stellar kinematics and the gravitational field on small scales. While future higher resolution cosmological simulations may partially alleviate this limitation, small-scale features in the predicted DM density field of the solar neighbourhood (Fig.~\ref{fig:gaia_pred}) should be interpreted with caution.

\subsection{Offset in density estimation} \label{sec:discuss2}

\begin{figure*}
    \centering
    \includegraphics[width=\textwidth]{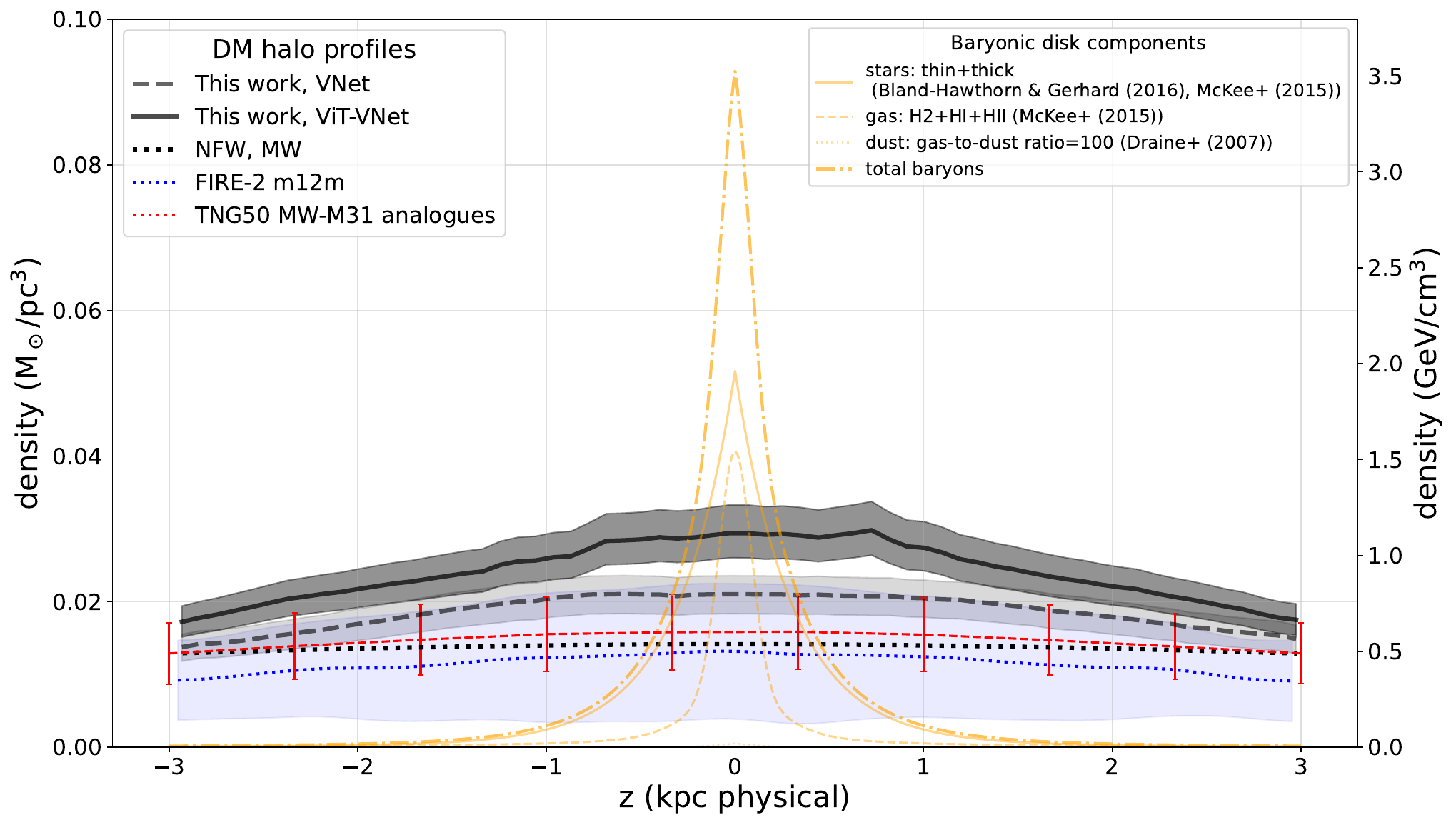}
    \caption{Vertical DM density profile at the solar radius. The grey dashed and solid lines show the predicted profiles from VNet and ViT-VNet, respectively, with shaded regions denoting test-based prediction deviations. The red and blue dotted lines show the profiles of the IllustrisTNG MW/M31 analogues and FIRE-2 m12m, respectively, while the black dotted line corresponds to the MW NFW profile, assuming $\rm M_{vir} = 1.3\times10^{12} \rm M_{\odot}$. The yellow lines show the vertical density profiles of the MW baryonic components. The m12m curve and band show the median and 16th--84th percentiles across test sample profiles located within a radius of $5.2 - 11.2 \, \rm kpc$; the TNG50 points and error bars show azimuthally averaged solar-radius profiles and the standard deviation among MW/M31 analogues.}
    \label{fig:zprofile}
\end{figure*}

Building on the reconstructed 3D DM density field (Sec.~\ref{sec:map}) and the scalar estimate of the local DM density (Sec.~\ref{sec:scala}), we further compare our results with theoretical halo models by examining the vertical DM density profile at the solar radius ($R = 8.122 \, \mathrm{kpc}$), as shown in Fig.~\ref{fig:zprofile}. The vertical profile is compared with the canonical NFW profile \citep{Navarro1997}, adopting MW parameters from \citet{BlandHawthorn2016}, as well as with profiles extracted from MW/M31-like systems in the TNG50 suite \citep{Pillepich2024} and one simulated galaxy from the FIRE-2 suite. For context, we also show representative MW baryonic components in the same figure, modelling each component using parameters from the literature \citep{Draine2007, McKee2015, BlandHawthorn2016}. The Gaia profiles are extracted from the reconstructed solar-neighbourhood volume, with additional downsampling applied to match the sampling inferred from the test samples. The m12m reference shows the median of the profiles from its held-out test volumes, together with the 16th--84th percentile range. The TNG50 reference uses azimuthal averages in a solar-radius ring for the MW/M31 analogues, with error bars indicating the standard deviation among the analogues.

The predicted vertical density profiles share a broad gradient with the simulation references, but their shapes are not identical: ViT-VNet shows a stronger disc-associated enhancement near the mid-plane. This distinct feature depends on the architecture, and edge effects should not be interpreted as established DM substructure. The overall normalisation is systematically higher than the reference profiles, by a factor of approximately $1.5-2$. This behaviour is consistently observed in both VNet and ViT-VNet predictions, with the latter yielding slightly higher amplitudes across the vertical range considered. The difference between the two models is larger than the test-based prediction deviations, indicating that the inferred density normalisation is subject to a model-dependent systematic uncertainty arising from both the architecture and the loss function. At the same time, this comparison should be interpreted in light of the domain shift between training and application: the consistent test performance of the two architectures shown in Fig.~\ref{fig:train_vis} and Fig.~\ref{fig:train_v2v} is evaluated against FIRE-2 ground truth within the simulation domain, whereas applying the trained models to Gaia remains an out-of-distribution task, even after matching the kinematic distributions and spatial sampling.

The normalisation offset shared by both architectures may therefore reflect residual differences between the dynamical state of the FIRE-2 galaxies used for training and that of the MW, in addition to architecture-dependent effects. The selection effects discussed in Sec.~\ref{sec:method_sel} may also contribute through their influence on the voxelwise mean velocities supplied to the networks.

We first attempted to mitigate these selection effects through kinematic matching by selecting simulated stellar particles whose kinematics resemble those of the Gaia sample (see Appendix~\ref{app:kinematics}). The approaches included isolating thin disc components using $\alpha$-element abundances, velocity phase-space diagrams and stellar particle mass, as well as applying the overall kinematic domain matching described above.  However, such subsampling cannot fully reproduce the intrinsic dynamical state of the MW. In particular, the rejection sampling based on $P_{\rm acc}$ matches only the marginal one-dimensional velocity distributions. The voxel-averaged velocity fields used as model inputs are instead determined by the joint distribution and spatial coherence of the observable velocities. Therefore, marginal velocity matching reduces part of the domain shift, but does not guarantee that higher-order correlations between stellar kinematics and the underlying gravitational potential are also matched.

The controlled sampling tests in Appendix~\ref{app:counts} show that downsampling Gaia towards the simulated occupancy lowers the inferred density for both architectures. Velocity-only inputs can therefore carry tracer-density information through noise, smoothness and empty-voxel patterns.

These limitations are further complicated by the selection effects in the Gaia sample itself. The Gaia sample is defined in observational space, including the magnitude cuts used to construct the selected sample, whereas the FIRE-2 domain matching is performed in velocity space. These selections therefore do not operate in the same variable space. Considered alongside the reported kinematic peculiarity of the MW relative to the FIRE-2 training set \citep{McCluskey2025}, these tests identify sensitivities to both sampling and kinematic matching. However, they do not show that the MW's kinematic coldness alone causes the offset or establish a unique downsampling correction.

Finally, both architectures were trained to infer the DM density field directly, rather than the total matter density field that determines the gravitational potential traced by stellar kinematics. The velocity-rescaling test in Appendix~\ref{app:beta} shows a systematic density response to changes in the input velocity amplitude, although the response is substantially ($\sim 2$ per cent for a $20$ per cent scaling in velocities) weaker than $\beta^2$. Together with the sampling tests discussed above, this suggests that the models potentially capture a connection between stellar kinematics and density while also learning correlations associated with tracer sampling. A more physically motivated approach would be to reconstruct the total density field first and then subtract the baryonic components to obtain the DM density. However, such a procedure would introduce additional uncertainty from baryonic subtraction, especially in the disc region where baryons can dominate over DM (see Fig.~\ref{fig:zprofile}). In the current framework, the baryon-to-DM relation of the FIRE-2 training galaxies is instead implicitly encoded in the simulation prior, making the inferred normalisation sensitive to differences in this relation between FIRE-2 and the MW. Further tests of tracer sampling, kinematic matching and distance selection, together with validation across additional galaxies, are needed to quantify these systematic effects and calibrate the absolute density normalisation.

\section {Summary and conclusion} \label{sec:concl}

Throughout this study, we explored whether the local DM density field in the solar neighbourhood can be inferred directly from stellar phase-space information using DL techniques. In contrast to conventional dynamical modelling that assumes equilibrium and specific parametric halo profiles, our framework learns this relation directly from simulations in a data-driven manner, without imposing an explicit analytic form of the gravitational potential. We trained and tested two DL architectures, VNet and ViT-VNet, to reconstruct the DM density field within a $(6\,\mathrm{kpc})^3$ volume centred on the solar position.

\begin{itemize}
    \item We trained and tested two distinct architectures, VNet and ViT-VNet, using volumetric samples constructed from $7$ MW-like galaxies in the FIRE-2 cosmological hydrodynamical simulation suite. Stellar particles were subsampled to match the kinematic properties of the Gaia sample. $9240$ volumetric cubes were generated for training, validation, and testing.
    \item While the input tensors consist of the three-dimensional velocity components ($v_{\alpha*}$, $v_{\delta}$, and RV), and the output tensor corresponds to a single three-dimensional DM density field defined on the same voxel grid, each architecture was trained $10$ times with randomised initialisation. We report the held-out prediction deviations which incorporate sensitivity to random initialisation, the Gaia ensemble mean, and architecture dependence separately.
    \item After training, the large-scale DM density gradient is well reproduced by both architectures, while small-scale density fluctuations are less accurately recovered, likely due to simulation resolution limits and the integrated nature of stellar velocity information.
    \item Gaia DR3 data were used to construct the input kinematic fields for the solar neighbourhood. An absolute magnitude cut of $M_G < 3.22$ was applied after extinction correction to reduce distance-dependent selection.
    \item The best estimates of local DM densities are $0.80^{+0.10}_{-0.10}$ and $1.12^{+0.15}_{-0.13}\,\mathrm{GeV}\,\mathrm{cm}^{-3}$ from VNet and ViT-VNet, respectively. The corresponding uncertainties reflect the inference scatter across the training set. These estimates are sensitive to tracer sampling and kinematic domain matching, suggesting the need for further investigation of their absolute calibration. Both models predict broad vertical gradients with a systematic upward shift relative to the reference profiles; ViT-VNet additionally shows a stronger near-disc enhancement. This common offset may further reflect a domain mismatch between the training set from FIRE-2 simulation and the MW. The difference between the two inferred values suggests a model-dependent systematic uncertainty.
\end{itemize}

Overall, both independent architectures converge to qualitatively consistent large-scale density structures during training, despite model-dependent differences in the inferred normalisation. This suggests that the models can extract information from stellar kinematics to constrain the global DM density gradient under the FIRE-2 simulation prior. However, the absolute normalisation remains sensitive to domain mismatch between the training simulations and the MW application. In particular, the MW has been identified as a kinematic outlier relative to the FIRE-2 training set, and the Gaia selection function and survey depth limit how well the observed sample represents the spatial and kinematic distributions of the full stellar population. We confirmed that these effects contribute to the systematic shift in the inferred density normalisation. Therefore, the absolute density normalisation requires further investigation; the present results demonstrate the mapping method and its systematic sensitivities.

The systematics identified in this work suggest several paths for future improvement. First, extending the training set to include other simulation suites, such as TNG50 \citep{Pillepich2024}, EAGLE \citep{Schaye2015}, or Auriga \citep{Grand2017}, would also help assess the dependence of the inferred density field on the adopted simulation prior. Second, the input representation can be expanded with additional voxel-level statistics, such as the velocity dispersion field, which can be computed directly from the same stellar particles used to construct the mean velocity field. Third, the kinematic matching procedure can be improved beyond marginal one-dimensional velocity matching by incorporating joint or conditional distributions, or by using importance-weighted training to account for differences between the simulated and observed samples. Fourth, future time-domain astrometric data from Gaia DR4 and cadence surveys such as LSST \citep{Ivezic2019} may enable stellar acceleration fields to be used as additional dynamical inputs. Finally, a more physically explicit route would be to infer the total matter density field and subtract baryonic components using an external baryonic model or prior, rather than directly predicting the DM density alone.

We anticipate that this methodology provides a complementary avenue for estimating the local DM density alongside traditional dynamical modelling approaches after proper calibration. The forward paths outlined above provide a roadmap for reducing the main systematic uncertainties identified in this work, including simulation-prior dependence, kinematic domain mismatch, limited input representations, and baryon--DM separation. More flexible architectures tailored to stellar phase-space data, including point-cloud-to-image mapping approaches, remain a promising direction, although their full implementation was beyond the computational scope of the present study. Together with deeper stellar catalogues, higher-resolution cosmological simulations, and future time-domain astrometric measurements, these developments will be essential for tightening the connection between stellar kinematics and the underlying DM density field.

\section*{Acknowledgements}
We thank the referees for the constructive comments, which have improved the manuscript.
YK, HSH, and JHK are supported by the Global-LAMP Program of the National Research Foundation of Korea (NRF) grant funded by the Ministry of Education (No. RS-2023-00301976).
HSH also acknowledges support from the National Research Foundation of Korea (NRF) grant funded by the Korea government (MSIT; RS-2026-25482692).

DJ was supported by NSF grant AST-2407298 at PSU, as well as by a KIAS Individual Grant, PG088301.

SEH and MY are supported by the Korea Astronomy and Space Science Institute grant funded by the Korea government (KASA, Korea AeroSpace Administration, 2026183201)

JHK’s work was also supported by the National Research Foundation of Korea (NRF) grant funded by the Korea government (MSIT) (No. 2022M3K3A1093827 and No. 2023R1A2C1003244).  His work was also supported by the National Institute of Supercomputing and Network/Korea Institute of Science and Technology Information with supercomputing resources including technical support, grants KSC-2024-CRE-0232. 

This work was also supported by the Korean Astronomy Machine Learning(KAML) working group.

\section{Data Availability}

The model architectures, trained model weights, training procedures, sample-generation pipeline, and Gaia DR3 data selection and retrieval procedures will be made publicly available following the completion of further work incorporating Gaia DR4.



\FloatBarrier
\bibliographystyle{mnras}
\bibliography{main} 




\appendix

\section{Ablation test for input channel selection}\label{sec:abltest}

\begin{figure*}
    \centering
    \includegraphics[width=170mm]{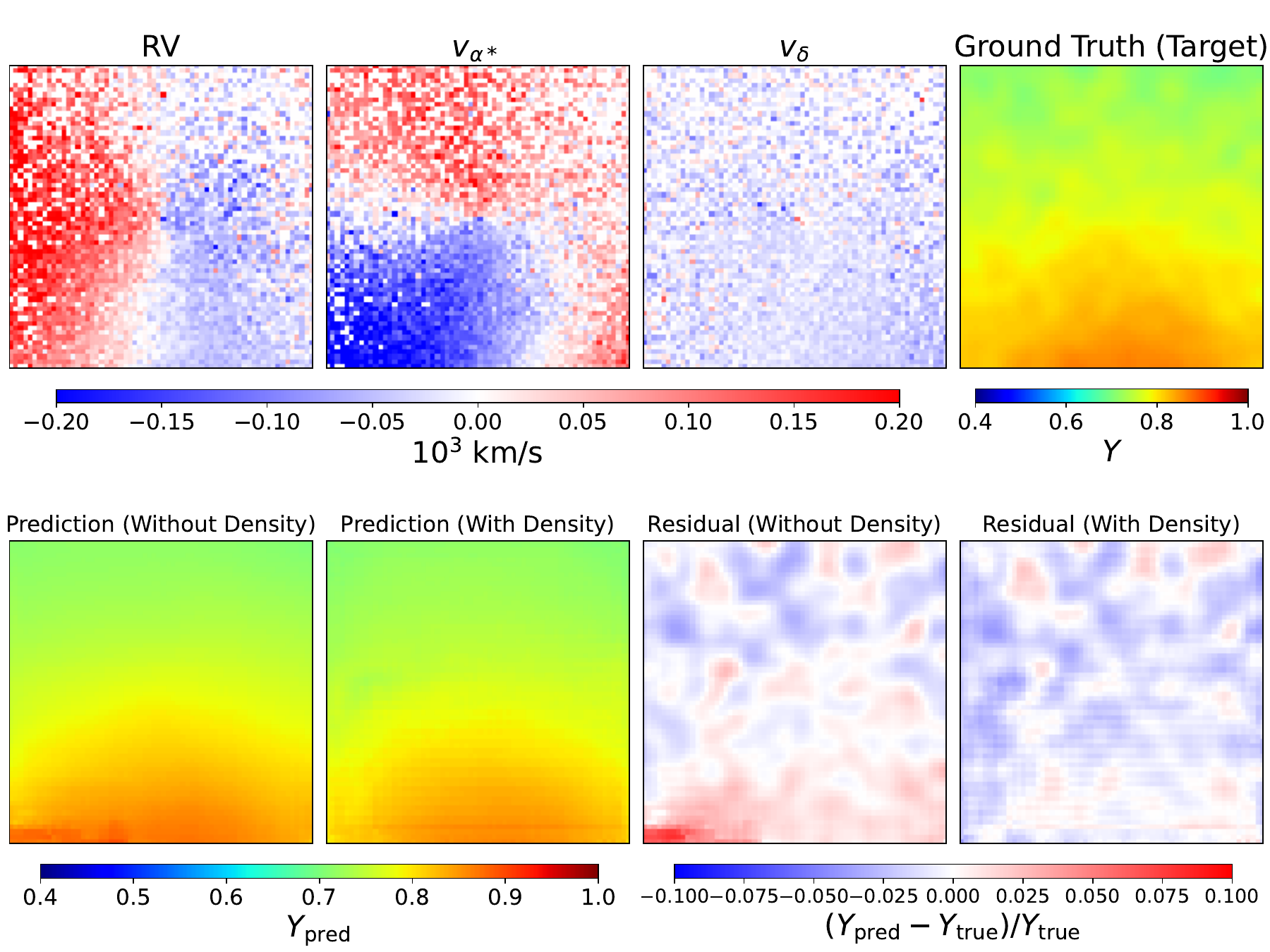}
    \caption{Same as Fig.~\ref{fig:train_vis}, but the results are from an identical VNet architecture presenting different test results from training with and without the stellar density channel. The density prediction and its residual obtained from the model trained without the stellar density channel are noted as ``Without Density" in the panel titles, while the result with the stellar density channel is noted as ``With Density".}
    \label{fig:train_vis_ablation}
\end{figure*}

As noted in Sec.~\ref{sec:method_sel}, obtaining a completely spatially complete star sample from observations equivalent to that in the simulation is limited due to extinction and the depth of the survey. We conducted an ablation test to investigate how excluding the stellar density channel affects the model performance, with the VNet architecture. All of the training setup, including the loss function, optimiser, batch size, learning rate and learning rate scheduler, is identical, but only the input has one additional channel corresponding to the stellar density channel.

Table~\ref{tab:density_ablation} shows the statistical scores of the test results from the two models. We also present the visualised results from a single test sample in Fig.~\ref{fig:train_vis_ablation}. The test sample is identical to that used in Fig.~\ref{fig:train_vis} for a fair comparison of DM density inference with and without the density channel. We confirmed that model performance was enhanced by adding the stellar density channel to the input. The improvement is small but statistically resolved by the quoted bootstrap intervals. The added mean signed bias in Table~\ref{tab:density_ablation} measures the test-set average offset.

\begin{table}
\centering
\begin{tabular}{ccc}
\toprule
Metric & With stellar density & Without stellar density \\
\midrule
$R^2$ & $0.91^{+0.002}_{-0.003}$ & $0.88^{+0.001}_{-0.001}$ \\
\\
Pearson $r$ & $0.96^{+0.001}_{-0.001}$ & $0.94^{+0.0005}_{-0.0006}$ \\
\\
nRMSE & $0.033^{+0.002}_{-0.0004}$ & $0.040^{+0.002}_{-0.0001}$ \\
\\
Bias & $0.00040^{+0.00003}_{-0.00003}$ & $-0.00060^{+0.00004}_{-0.00004}$ \\
\bottomrule
\end{tabular}
\caption{Ablation-test results for the VNet model with and without the stellar number density field as an additional input channel. The reported values are evaluated on the test set. Metrics use the transformed density $Y$; bias is $\langle Y_{\rm pred}-Y_{\rm true}\rangle$ and nRMSE is RMSE divided by $\max(Y_{\rm true})-\min(Y_{\rm true})$. Reported confidence intervals correspond to the $68$ per cent range, equivalent to the $1\sigma$ uncertainty estimated via bootstrap resampling.}
\label{tab:density_ablation}
\end{table}

\section{Kinematic domain matching and number-density mismatch}
\label{app:B}

In this appendix, we examine the kinematic domain-matching procedure and the mismatch in tracer number density between FIRE-2 and Gaia. These tests quantify sensitivity of the reconstruction rather than provide a complete observational calibration.

\subsection{Effect of sample selection and kinematic domain matching}\label{app:kinematics}

\begin{figure}
\centering
\includegraphics[width=\columnwidth]{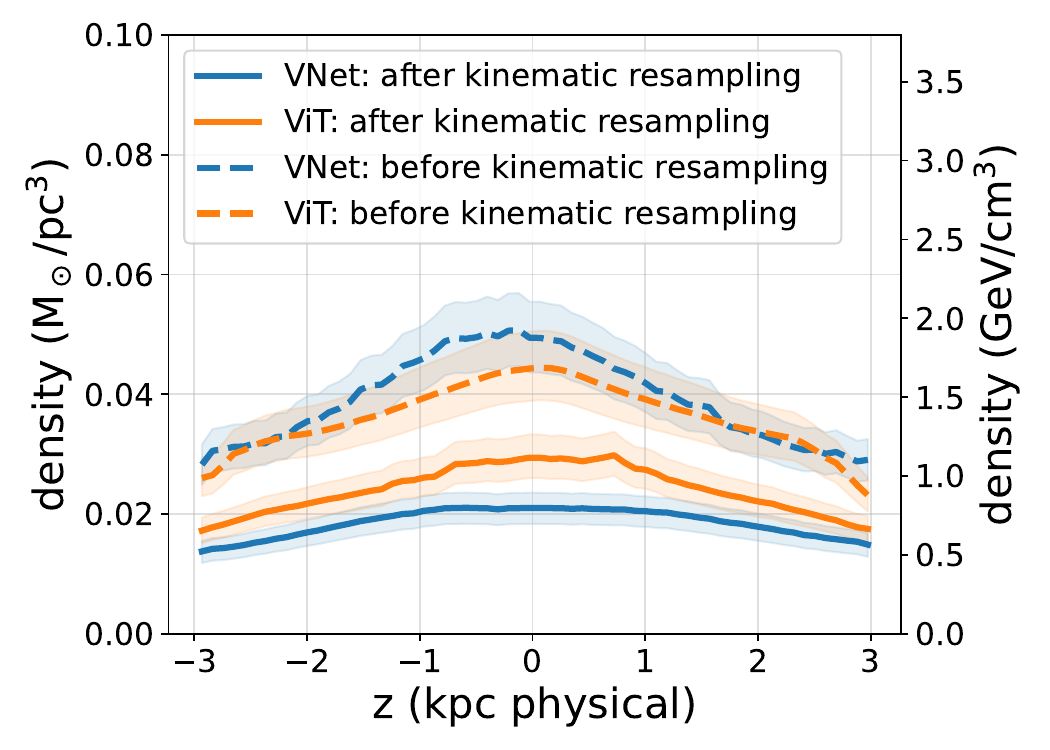}
\caption{Same as Fig.~\ref{fig:zprofile}, but comparing the vertical DM density profiles inferred from models trained before and after applying the rejection-sampling procedure. The dashed lines show the predictions from models trained without rejection sampling, while the solid lines show the main results obtained after kinematic domain matching.}
\label{fig:zprofile_sel}
\end{figure}

We first investigate how the rejection sampling applied to FIRE-2 stellar particles affects the inference. This procedure improves the consistency between the marginal velocity distributions of the Gaia sample and the FIRE-2 training inputs, but it should not be interpreted as a complete validation of the baryon--DM relation after selection. In particular, matching the one-dimensional velocity distributions does not guarantee that the joint velocity distribution, spatial coherence of the voxel-averaged velocity field, or the correlation between stellar kinematics and the underlying gravitational potential are also matched. The rejection sampling may therefore modify the effective relation between the selected stellar population and the DM density field.

To quantify the impact of this procedure, Fig.~\ref{fig:zprofile_sel} compares the vertical DM density profiles inferred using models trained before and after applying the rejection sampling. The domain-matched training set shifts the inferred density normalisation towards lower values. This suggests that the kinematic matching might reduce part of the domain mismatch between FIRE-2 and Gaia. Nevertheless, the remaining offset discussed in Sec.~\ref{sec:discuss2} indicates that the rejection sampling does not fully remove differences between the MW and the simulated training galaxies.

A more stringent validation would require applying an observationally motivated Gaia selection function to mock Gaia catalogues constructed from the same simulations, for example using the Ananke framework \citep{Sanderson2020, Nguyen2024_Ananke, CantatGaudin2023, CastroGinard2023}, and then testing whether the corresponding DM density field can be recovered. We did not perform such a mock-catalogue validation in this work. Therefore, the rejection-sampling procedure should be regarded as a practical first-order domain matching step rather than a fully validated correction for Gaia selection effects or baryon--DM relation changes.

\subsection{Tracer number density and downsampling}\label{app:counts}

\begin{figure}
\centering
\includegraphics[width=\columnwidth]{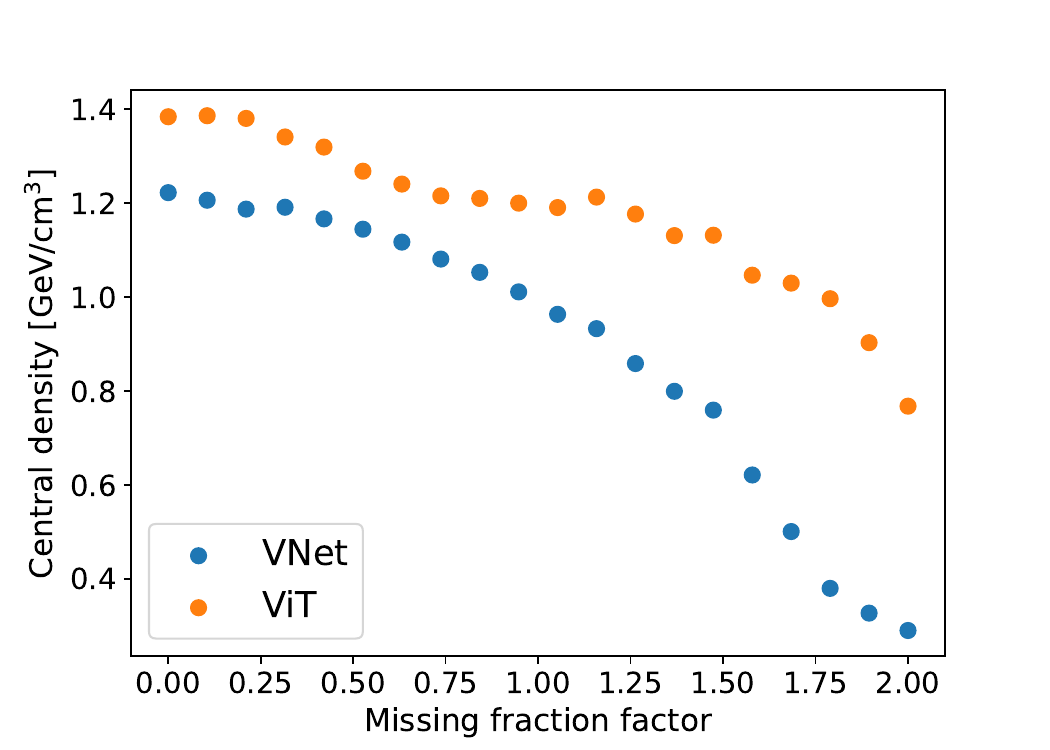}
\caption{Central-eight-voxel density predictions versus the factor $h$ multiplying the m12m-based vertical downsampling prescription, with a fixed trained realisation of each architecture. Both models show a systematic decrease in normalisation as downsampling increases.}
\label{fig:missingness}
\end{figure}

Although the stellar number-density channel is excluded, tracer counts affect the noise, smoothness and empty-voxel patterns of the input velocity fields. Keeping the trained models fixed, interpolation of empty m12m input voxels increases the inferred density normalisation, while downsampling the Gaia inputs towards the m12m occupancy lowers it. The reference vertical downsampling pattern is derived from the m12m test samples. The factor $h$ in Fig.~\ref{fig:missingness} multiplies that reference pattern: $h=0$ applies no additional downsampling and $h=1$ uses the reference prescription.

The broad reconstructed gradient can remain similar while the normalisation changes. The scan demonstrates sampling sensitivity, but does not determine the true MW density or a uniquely justified correction. Although we retain the input downsampled by a factor of $h=1$ for the main Gaia results, we note that the inferred normalisation is sensitive to the spatial completeness of the input channels.

\section{Response to velocity rescaling}\label{app:beta}

\begin{figure}
\centering
\includegraphics[width=\columnwidth]{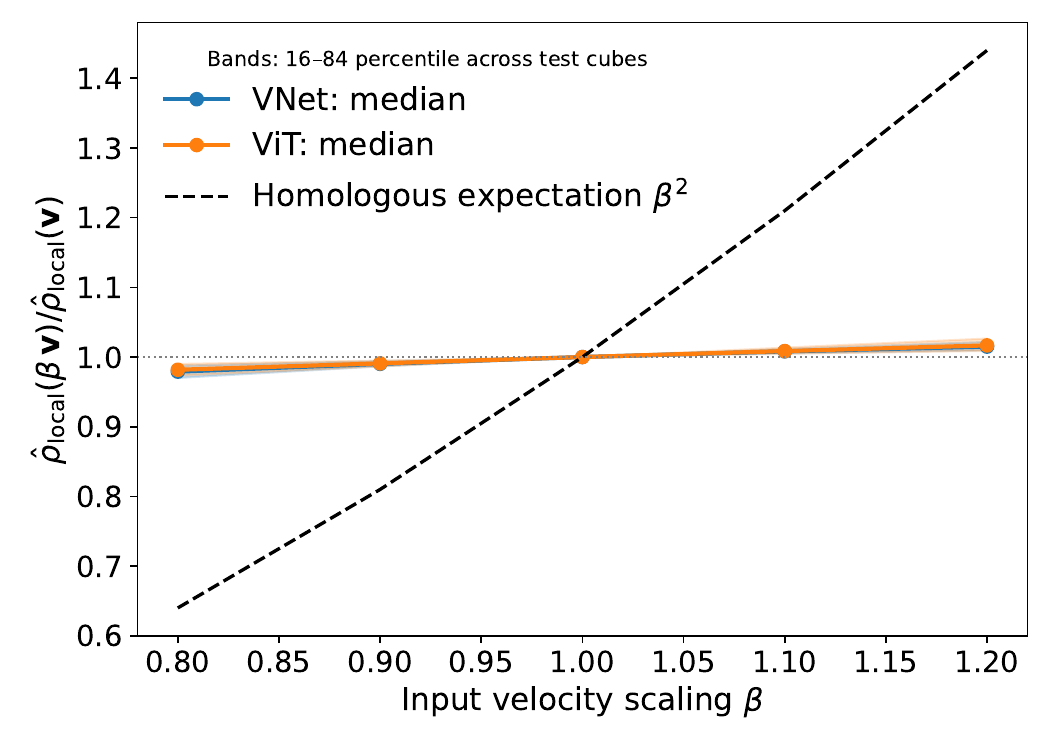}
\caption{Response of the central-eight-voxel mean prediction to input-velocity rescaling. The line and band show the median and 16th--84th percentiles across m12m test cubes for a fixed trained model. The dashed line is the $\beta^2$ reference relation.}
\label{fig:beta_central}
\end{figure}

To test whether the model learned the actual physical relationship between tracer kinematics and the density field, we multiply the m12m input velocities by $\beta=0.8$--$1.2$ at fixed positions and apply the same trained model without retraining. Figure~\ref{fig:beta_central} shows the ratio of the mean prediction in the eight central voxels to its value at $\beta=1$, with the median and 16th--84th percentiles across test cubes.

The predicted density increases systematically with velocity amplitude, but the response ($\sim 2$ per cent for a $20$ per cent scaling in velocities) is substantially weaker than $\beta^2$. A self-consistent steady-state rescaling would require $\Phi\rightarrow\beta^2\Phi$ and $\rho_{\rm tot}\rightarrow\beta^2\rho_{\rm tot}$. This diagnostic instead changes only the input velocities of a fixed model predicting DM alone, leaving the baryonic field and physical target unchanged. The response demonstrates a kinematic dependence and a limitation of its dynamical calibration, rather than either complete insensitivity to motion or recovery of a general mass-scaling law.


\bsp	
\label{lastpage}
\end{document}